\documentclass[aps,prd,notitlepage,nofootinbib,preprintnumbers,superscriptaddress]{revtex4}
\usepackage{amssymb}
\usepackage{amsmath}
\usepackage{verbatim}
\usepackage{appendix}
\usepackage[usenames,dvipsnames]{xcolor}

\begin{document}

%Revised version
\preprint{YITP-15-49, IPMU15-0102}
\title{Derivative-dependent metric transformation and physical degrees of freedom}

\author{Guillem Dom\`enech}
%\email{guillem.domenech@yukawa.kyoto-u.ac.jp}
\affiliation{Yukawa Institute for Theoretical Physics, Kyoto University, 606-8502, Kyoto, Japan}

\author{Shinji Mukohyama}
%\email{shinji.mukohyama@yukawa.kyoto-u.ac.jp}
\affiliation{Yukawa Institute for Theoretical Physics, Kyoto University, 606-8502, Kyoto, Japan}
\affiliation{Kavli Institute for the Physics and Mathematics of the Universe (WPI), The University of Tokyo Institutes for Advanced Study, The University of Tokyo, Kashiwa, Chiba 277-8583, Japan}

\author{Ryo Namba}
%\email{ryo.namba@ipmu.jp}
\affiliation{Kavli Institute for the Physics and Mathematics of the Universe (WPI), The University of Tokyo Institutes for Advanced Study, The University of Tokyo, Kashiwa, Chiba 277-8583, Japan}

\author{Atsushi Naruko}
%\email{naruko@th.phys.titech.ac.jp}
\affiliation{Department of Physics, Tokyo Institute of Technology, %2-12-1 Ookayama, 
Meguro-ku, Tokyo 152-8551, Japan}

\author{Rio Saitou}
%\email{rio.saitou@ipmu.jp}
\affiliation{Institute for Cosmic Ray Research, University of Tokyo, Kashiwa, Chiba 277-8582, Japan}

\author{Yota Watanabe}
%\email{yota.watanabe@ipmu.jp}
\affiliation{Kavli Institute for the Physics and Mathematics of the Universe (WPI), The University of Tokyo Institutes for Advanced Study, The University of Tokyo, Kashiwa, Chiba 277-8583, Japan}

\begin{abstract}
We study metric transformations which depend on a scalar field $\phi$ and its first derivatives and confirm that the number of physical degrees of freedom does not change under such transformations, as long as they are not singular. We perform a Hamiltonian analysis of a simple model in the gauge $\phi = t$. In addition, we explicitly show that the transformation and the gauge fixing do commute in transforming the action. We then extend the analysis to more general gravitational theories and transformations in general gauges. We verify that the set of all constraints and the constraint algebra are left unchanged by such transformations and conclude that the number of degrees of freedom is not modified by a regular and invertible generic transformation among two metrics. We also discuss the implications on the recently called ``hidden" constraints and on the case of a singular transformation, a.k.a. mimetic gravity.
\end{abstract}

\maketitle

\section{Introduction}

Conformal and disformal metric transformations have been useful tools in studying various aspects of gravitational theories. A metric is simply a non-degenerate symmetric rank-$2$ tensor. Hence, one can easily construct a new metric by combining the original metric variable with other fields in the gravity sector. Picking up a particular metric is nothing but fixing a basis or a representation in the field space, provided that the transformation from the original basis is regular. For this reason a choice of a metric variable and a transformation among different choices is often called a frame and a frame transformation, respectively. In contrast, one might consider it more appropriate to use the word ``representation'' rather than ``frame'', in line with Dicke's original paper \cite{Dicke:1961gz} (see also \cite{Deruelle:2010ht} for a recent remark on this point). Nevertheless, here we adopt the common terminology in the literature, namely ``frame''. A conformal transformation is a special case of frame transformations: the new and the original metrics are conformally equivalent to each other. A disformal transformation is a frame transformation which does not fit in this criterion while, for the simple cases studied in the present paper and in most of the literature so far, it can be understood as a rescaling of the lapse function keeping the spatial metric unchanged under a certain gauge choice as we see later. For an alternative interpretation, a disformal transformation can be perturbatively thought of as a rescaling of time (see \cite{Domenech:2015hka}).

A frame transformation may change the structure of kinetic mixing between the metric and other fields in the gravity sector. The description of the gravity sector is often simplified by going to a frame where such kinetic mixing is minimized. If possible, it would thus be the best to choose a frame in which (the lowest dimensional part of) the kinetic term for the metric is simply the Einstein-Hilbert action. Such a frame is called Einstein frame, if exists at all in a given theory (for an analysis in the scalar-tensor theory see \cite{Bettoni:2013diz,bettoni2015kinetic}). On the other hand, the description of matter fields is simple in a frame where they couple to the metric minimally, i.e. a frame in which free particles made of matter fields follow geodesics. Such a frame is usually called Jordan frame or matter frame. It is thus rather important to understand how the description of a theory changes from one frame to the other, especially from the Einstein frame to the matter frame, and vice versa. Along this direction, the invariance of physical observables under conformal and disformal transformations is intensively studied in the literature \cite{Catena:2006bd,Faraoni:2006fx,Deruelle:2010ht,White:2012ya,White:2013ufa,Chiba:2013mha,Jarv:2014hma,Minamitsuji:2014waa,Creminelli:2014wna,Tsujikawa:2014uza,Watanabe:2015eia,Watanabe:2015uqa,motohashi2015dis,Domenech:2015hka}.

In the present paper we consider a class of frame transformations that are of the form 
\begin{equation}
 g_{\mu\nu} \to \tilde{g}_{\mu\nu} = \mathcal{A}(\phi,X) g_{\mu\nu} + \mathcal{B}(\phi,X)\partial_{\mu}\phi\partial_{\nu}\phi, \label{eqn:frametr}
\end{equation}
where $\phi$ is a scalar field and $X=-g^{\mu\nu}\partial_{\mu}\phi\partial_{\nu}\phi$. Concerning the first term on the right-hand side (so-called Weyl transformation part), we introduce a general conformal factor, ${\cal A}$, which can now depend on the first derivative of the field as well as the field itself. As for the second term, this type of frame transformations was originally introduced by Bekenstein~\cite{Bekenstein:1992pj}. It should be noted that ${\cal A}$ and ${\cal B}$ are such that by definition no change in the signature of the metric is allowed.

As an example of the systems that naturally involve frame transformations, let us consider a D$3$-brane moving in a higher dimensional spacetime. In the type IIB string theory, all complex structure moduli are stabilized by turning on various fluxes~\cite{Giddings:2001yu}. Those fluxes act as gravitational sources and, as a result, the geometry of the extra dimensions is inevitably warped. A warped region of the extra dimensions is called a warped throat and has a $10$-dimensional metric of the form 
\begin{equation}
 ds_{10}^2 = h^2g_{\mu\nu}d\xi^{\mu}d\xi^{\nu}+h^{-2}\gamma_{pq}d\xi^pd\xi^q,
  \label{eqn:10d-metric}
\end{equation}
where $g_{\mu\nu}$ ($\mu,\nu=0,\cdots,3$) is a $4$-dimensional metric, $\gamma_{pq}$ ($p,q=4,\cdots,9$) is a Calabi-Yau metric on a $6$-dimensional compact manifold and the warp factor $h$ depends only on the internal Calabi-Yau direction. If K\"ahler moduli are also stabilized, e.g. {\it \'a la} Kachru, Kallosh, Linde and Trivedi~\cite{Kachru:2003aw}, then the low-energy behavior of the metric $g_{\mu\nu}$ is described by the $4$-dimensional Einstein gravity. We can thus regard $g_{\mu\nu}$ as the Einstein-frame metric. The bosonic part of a probe $D3$-brane action is given as the sum of a Dirac-Born-Infeld (DBI) action, 
%============< EQUATION >==============%
%
\begin{equation}
 I_{\mathrm{DBI}} = -T_3\int d^4x
  \sqrt{-\det(\tilde{g}_{\mu\nu}+2\pi\alpha'F_{\mu\nu})} \; ,
\end{equation}
%======================================%
and a Chern-Simon action, where $x^{\mu}$ ($\mu=0,\cdots,3$) are the intrinsic coordinates on the brane, $T_3$ is the brane tension, $\tilde{g}_{\mu\nu}$ is the induced metric on the brane, $\alpha'$ is the Regge slope, and $F_{\mu\nu}$ is the field strength of the $U(1)$ gauge field on the brane. Here, for simplicity we have assumed that the pullback of the NS-NS antisymmetric field on the brane world-volume vanishes and that the dilaton is stabilized as in the construction by Giddings, Kachru and Polchinski~\cite{Giddings:2001yu}. Expanding the DBI action with respect to $F_{\mu\nu}$, it is easy to see that the $U(1)$ gauge field on the brane minimally couples to the induced metric $\tilde{g}_{\mu\nu}$ (instead of $g_{\mu\nu}$).\footnote{The Chern-Simon term does not depend on $F_{\mu\nu}$.} We can thus regard $\tilde{g}_{\mu\nu}$ as the matter (or Jordan) frame metric. We consider the warped geometry of the form (\ref{eqn:10d-metric}) with 
%============< EQUATION >==============%
%
\begin{equation}
 \gamma_{pq}d\xi^pd\xi^q = dr^2 + ds_5^2, \quad h = h(r), 
\end{equation}
%======================================%
where $ds_5^2$ is an $r$-dependent $5$-dimensional metric. For example the Klebanov-Strassler throat~\cite{Klebanov:2000hb} is indeed of this form. We consider the radial motion of a $D3$-brane in the warped geometry, adopting a gauge in which the four coordinates on the brane $x^{\mu}$ ($\mu=0,\cdots,3$) coincide with the $4$-dimensional part of the bulk coordinates $\xi^{\mu}$, and supposing that the brane radial position $r$ also depends on $x^{\mu}$. In this case, 
\begin{equation}
 \tilde{g}_{\mu\nu} = h(r)^2 g_{\mu\nu} + h(r)^{-2}\partial_{\mu}r\partial_{\nu}r. \label{eqn:example}
\end{equation}
Considering the brane position $r$ as a scalar field in the $4$-dimensions, one sees that this is a special case of the frame transformation (\ref{eqn:frametr}). The form of the DBI action is completely fixed by Lorentz invariance of the D$0$-brane action and T-duality. Hence, the DBI action and thus the frame transformation are valid even for a relatively large value of $\partial_{\mu}r$, if the signature of $\tilde{g}_{\mu\nu}$ remains the same and if the second (and higher) derivatives of $r$ (and the curvature) are sufficiently small in the unit of the string scale. In this example, however, the conformal and disformal factors $\mathcal{A}$ and $\mathcal{B}$ are simply functions of the scalar field and thus independent of its derivatives.

In the effective field theory approach, on the other hand, all possible terms should be included in the action as far as they are consistent with the symmetries and/or the symmetry breaking patterns that are invoked for the construction of the theory. For example, in the covariant formulation of ghost condensation~\cite{ArkaniHamed:2003uy}, terms nonlinear in $X$ are not suppressed compared with terms linear in or independent of $X$, but one can still construct a sensible low energy effective field theory. An extension of such a construction to an expanding universe leads to the effective field theory of inflation~\cite{Cheung:2007st}. In these effective field theories, one does not intend to expand a theory around $X=0$. Instead, one expands a theory around a non-vanishing background value of $X$. From this point of view, as far as we are interested in a background with non-vanishing $X$, it is rather natural to expect that nonlinear (as well as linear) $X$-dependence of the conformal and disformal factors $\mathcal{A}$ and $\mathcal{B}$ is not suppressed compared with constant parts, unless a specific UV completion realizes such suppression. On the other hand, dependence on higher derivatives is under control and can be safely ignored if we choose a background on which $X$ does not change as fast as $\phi$. For these reasons, in the present paper we maintain the general $X$-dependence of the conformal and disformal factors $\mathcal{A}$ and $\mathcal{B}$.

Under a certain condition, the frame transformation (\ref{eqn:frametr}) is invertible and thus relates two equivalent theories to each other. In the case where the conformal factor ${\cal A}$ is independent of $X$, the action transformed from the Einstein-Hilbert action through \eqref{eqn:frametr} falls into the class of theories proposed by Gleyzes et al.~\cite{Gleyzes:2014dya} (it reduces to a sub-class of the Horndeski theory \cite{Horndeski:1974wa,Deffayet:2011gz,Kobayashi:2011nu} when the disformal factor ${\cal B}$ also depends only on $\phi$). The number of degrees of freedom in this class of theories has been proven to be the same as that of a scalar plus general relativity (GR) at least up to the so-called ${\cal L}_4$ term in the so called unitary gauge $\phi=t$ (see (\ref{eqn:unitary-gauge}) and discussions thereafter) \cite{Lin:2014jga,Gleyzes:2014qga} (see also \cite{Gao:2014fra} for Hamiltonian analysis of theories proposed in \cite{Gao:2014soa}, as an extension of~\cite{Gleyzes:2014dya}). The proof for the ${\cal L}_4$ term of the generalized Galileon theory in the general gauge has been done recently in \cite{deffayet2015counting}. However, if ${\cal A}$ depends on $X$, transformations of the type \eqref{eqn:frametr} relate a theory with second-order equations of motion to that with higher-order equations of motion~\cite{Bettoni:2013diz}. Those higher-order field equations can be reduced to a set of second order differential equations in the unitary gauge but the appearance of the first and second time derivatives of the lapse function will spoil the nature of the Hamiltonian constraint. One might thus mistakenly conclude that the latter theory would have more physical degrees of freedom than the former. Of course this is not the case since the two theories are equivalent to each other as long as the transformation is not singular (see also \cite{Gleyzes:2014rba} where the healthiness, to be precise, the degeneracy of the kinetic matrix and thus the existence of a non-trivial constraint of such theories with derivatives of the lapse function was pointed out at the level of perturbation). In order to reconcile the apparent contradiction we apply the standard method of Hamiltonian analysis for a constrained system. We then conclude that frame transformations of the form (\ref{eqn:frametr}) indeed do not change the number of physical degrees of freedom, provided that the transformation is regular and invertible.

The rest of the present paper is organized as follows. In Sec.~\ref{sec:frametr}, as a simple example, we disformally transform the Einstein-Hilbert action by \eqref{eqn:frametr} without fixing any gauge and show that the terms with second time derivatives of $\phi$ cannot be eliminated from the action in general. We also derive the action in the unitary gauge and show the presence of terms quadratic in $\dot N$ ($N$ is the lapse function) provided that ${\cal A}$ depends on $X$. In Sec.~\ref{sec:analysis}, we apply the Hamiltonian analysis for a constrained system to the transformed action in the unitary gauge and prove that the presence of nontrivial constraints keeps the number of physical degrees of freedom unchanged as compared to the original action. In Sec.~\ref{sec:general}, we generalize our proof to more general gravitational theories and transformations in general gauge. Viewing a transformation as a change of variables, we show that the constraint structure does not change, and the system evolves in the same way as that before the transformation. This proves that the number of physical degrees of freedom does not change. Finally in Sec.~\ref{sec:summary}, we summarize our result and discuss its implications. In Appendix \ref{sec:detailed}, we collect some details in the calculations of disformal transformation and of the gauge fixing. In Appendix \ref{app:example}, we show a simple example to systematically replace derivatives with auxiliary variables in any given transformation, with which the general analysis described in Sec.~\ref{sec:general} is applicable, and also point out that this procedure is not valid for singular transformations.

\section{Frame transformation}
\label{sec:frametr}

Let us consider a generic disformal transformation of the metric $\tilde g$ as in (\ref{eqn:frametr}) in an $d+1$ dimensional spacetime, where ${\cal A}$ and ${\cal B}$ are arbitrary functions of $\phi$ and $X\equiv -g^{\mu\nu}\partial_{\mu}\phi\, \partial_{\nu}\phi$ $(\mu =0\,,1\,,\cdots\,,d)$ (see also \cite{Kothawala:2014oba}). The inverse of the metric \eqref{eqn:frametr} is given by
\begin{equation}
\tilde g^{\mu\nu} = \frac{1}{{\cal A}} \left( g^{\mu\nu} - \frac{{\cal B}}{ {\cal A} - {\cal B}X} \, \nabla^\mu \phi \, \nabla^\nu \phi \right)\,,
\label{trans_inverse}
\end{equation}
provided that ${\cal A} ({\cal A} - {\cal B} X) \neq 0$,
where $\nabla_\mu$ is the covariant derivative with respect to $g_{\mu\nu}$. 
The inverse transformation
\begin{equation}
g_{\mu\nu} = \tilde {\cal A}(\phi,\tilde X) \, \tilde g_{\mu\nu} + \tilde {\cal B}(\phi,\tilde X) \, \partial_\mu \phi \, \partial_\nu \phi \; , \quad
 \label{eqn:inverse-transformation}
\end{equation}
where $\tilde X \equiv - \tilde g^{\mu\nu} \partial_\mu \phi \, \partial_\nu \phi$, is non-linearly related to the original one \eqref{eqn:frametr} through
\begin{equation}
\tilde {\cal A}\left( \phi , \frac{X}{{\cal A}-{\cal B}X} \right) = \frac{1}{{\cal A}(\phi , X)} \; , \quad
\tilde {\cal B} \left( \phi , \frac{X}{{\cal A}-{\cal B}X} \right) = - \frac{{\cal B}(\phi,X)}{{\cal A}(\phi,X)} \; ,
\end{equation}
provided that ${\cal A}({\cal A}-{\cal B}X)\neq 0$ and the Jacobian of the transformation is non-vanishing,
\begin{equation}
{\cal A}({\cal A}-{\cal A}_XX+{\cal B}_XX^2)\neq 0 \; ,
 \label{eqn:invertibility-condition}
\end{equation}
where ${\cal A}_X \equiv (\partial {\cal A} / \partial X) \, |_\phi$, ${\cal B}_X \equiv (\partial {\cal B} / \partial X) \, |_\phi$ \cite{Zumalacarregui:2013pma}.

We now perform the transformation \eqref{eqn:frametr} to the Einstein-Hilbert action, that is
\begin{equation}
\tilde I_{\rm EH} = \frac{M_p^2}{2} \int d^{d+1} x \sqrt{- \tilde g} \, \tilde R
 \,.
\label{EHaction}
\end{equation}
A straightforward calculation leads to the transformed action, given as (see Appendix \ref{sec:full} for details) 
\begin{align}
\tilde I_{\rm EH} & = \frac{M_p^2}{2} \int d^{d+1}x \sqrt{-g} \, {\cal A}^{(d-1)/2} \left( 1 - \frac{{\cal B}}{{\cal A}} X \right)^{1/2} 
\nonumber\\ & \qquad\qquad
\times \Bigg\{ R
 + \frac{{\cal B}}{{\cal A}-{\cal B}X}
 \Bigl[ \nabla_\mu \nabla_\nu \phi \, \nabla^\mu \nabla^\nu \phi - \left( \nabla^2 \phi \right)^2 \Bigr]
+ \frac{d(d-1)}{4} \Bigl[ \nabla_\mu \ln {\cal A} \, \nabla^\mu \ln {\cal A}
 - \frac{{\cal B}}{{\cal A}-{\cal B}X}
 \left( \nabla^\mu \phi \, \nabla_\mu \ln {\cal A} \right)^2 \Bigr]
\nonumber\\ & \qquad\qquad \qquad
- \frac{d-1}{2} \, \frac{{\cal B}}{{\cal A}-{\cal B}X} \, \nabla_\mu \ln {\cal A}
\left[ X \nabla^\mu \ln \left( \frac{\cal B}{\cal A} \right)
 + \nabla^\mu \phi \, \nabla^\nu \phi \, \nabla_\nu \ln \left( \frac{\cal B}{\cal A} \right) \right]
\nonumber\\ & \qquad\qquad \qquad
- \frac{{\cal B}}{{\cal A}-{\cal B}X}
 \left( \frac{1}{2} \nabla^\mu X + \nabla^\mu \phi \, \nabla^2 \phi \right) \left[ \left( d-1 \right) \nabla_\mu \ln {\cal A}
 + \nabla_\mu \ln \left( \frac{\cal B}{\cal A} \right) \right] \Bigg\} \; ,
\label{act_trans_full}	
\end{align}
up to total derivatives.  We shall later add a $k$-essence type action for the scalar field, but at this moment we first apply the transformation to the Einstein-Hilbert part.

It is worth noting that, in the case of $d+1 = 4$, for ${\cal A}={\cal A}(\phi)$ and ${\cal B}={\cal B}(\phi)$, the action \eqref{act_trans_full} reduces to a subclass of the Horndeski one \cite{Horndeski:1974wa,Deffayet:2011gz,Kobayashi:2011nu}, which has been known as a scalar-tensor theory with only $3$ degrees of freedom, concretely $2$ tensor and $1$ scalar. The crucial point is whether ${\cal A}$ and ${\cal B}$ depend on $X$, and if at least one of them does, the transformed action \eqref{act_trans_full} is not within the domain of the Horndeski theory. In fact, in the case ${\cal A}={\cal A}(\phi)$ and ${\cal B}={\cal B}(\phi,X)$ as discussed in \cite{Gleyzes:2014qga}, the action \eqref{act_trans_full} can be written as a subclass of the theory proposed by Gleyzes et al.~\cite{Gleyzes:2014dya} which has been shown to have no extra degrees of freedom at least in the unitary gauge (see (\ref{eqn:unitary-gauge}) and discussions thereafter for the definition of the unitary gauge)~\cite{Lin:2014jga,Gleyzes:2014qga}. When ${\cal A}$ depends on $X$, however, the terms quadratic in the second time derivatives of $\phi$, i.e.~such terms as proportional to $\nabla_\mu X \nabla^\mu X$, enter the action in a nontrivial manner \cite{Zumalacarregui:2013pma}. It is generically impossible to eliminate these terms by diagonalizing the kinetic matrix and this may at a first glance appear to introduce a ghost-like extra degree of freedom. However, as is clear from the derivation of the action \eqref{act_trans_full} from the original Einstein-Hilbert action \eqref{EHaction}, this is merely an artifact of the frame transformation \eqref{eqn:frametr} and the number of degrees of freedom should stay the same as long as the transformation is non-singular. In the next section we will explicitly show this claim by the Hamiltonian analysis for a simple case in the unitary gauge.

Before proceeding, in order to simplify the analysis, we fix the time gauge degree of freedom or time slicing in the action \eqref{act_trans_full}, while keeping the spatial gauge arbitrary, by taking
\begin{equation}
\phi = t \; , \label{eqn:unitary-gauge}
\end{equation}
where we have implicitly assumed the existence of such time-slicings, namely that $\partial_\mu \phi$ is non-zero, regular and timelike everywhere in both frames. We adopt the nomenclature in the literature and call this gauge choice ``unitary gauge.'' In this gauge, it is convenient to do the $(d+1)$-decomposition {\it \`{a} la} Arnowitt-Deser-Misner (ADM) as
\begin{equation}
 ds^2 = 
 g_{\mu\nu}dx^{\mu}dx^{\nu} = - N^2 dt^2 + \gamma_{ij} \left( dx^i + N^i dt \right) \left( dx^j + N^j dt \right) \; , \qquad 
 (i = 1 \,, 2 \,, \cdots \,, d)
 \label{eqn:ADM-g}
\end{equation} 
where $N$ is the lapse function, $N^i$ the shift vector, and $\gamma_{ij}$ the $d$-dimensional spatial metric. Spatial indices are lowered and raised by the spatial metric $\gamma_{i j}$ and its inverse $\gamma^{ij}$, respectively. With this decomposition, $X$ is simply expressed in this gauge as 
\begin{align}
 X = - g^{\mu \nu} \partial_\mu \phi \, \partial_\nu \phi = \frac{1}{N^2} \,.
\end{align}
And hence ${\cal A}$ and ${\cal B}$ are now understood as functions of $t$ and $N$ rather than $\phi$ and $X$. By adopting the ADM decomposition for $\tilde{g}_{\mu\nu}$ as well 
\begin{equation}
 d \tilde{s}^2
 = \tilde{g}_{\mu\nu}dx^{\mu}dx^{\nu} = -\tilde{N}^2 dt^2 + \tilde{\gamma}_{ij}(dx^i+\tilde{N}^idt)(dx^j+\tilde{N}^jdt) \; ,
 \label{eqn:ADM-gtil}
\end{equation} 
the transformation (\ref{eqn:frametr}) is rephrased as 
\begin{equation}
 N \to \tilde{N}, \quad N^i \to \tilde{N}^i, \quad
  \gamma_{ij} \to \tilde{\gamma}_{ij}, 
\end{equation}
where
\begin{equation}
 \tilde{N}^2 = {\cal A} N^2 - {\cal B}, \quad \tilde{N}^i = N^i, \quad
  \tilde{\gamma}_{ij} = {\cal A} \gamma_{ij}. \label{eqn:unitary-gauge-transformation}
\end{equation}
Now the physical meaning of a disformal transformation is clear. While a conformal transformation with ${\cal A}\ne 1$ (${\cal A} \neq 0$) and $\mathcal{B}=0$ scales both the lapse function and the spatial metric, a disformal transformation with $\mathcal{A}=1$ and $\mathcal{B}\ne 0$ induces a mere change of the lapse function. Such transformation does not involve time derivatives of the fields and is invertible, provided that 
\begin{equation}
 \frac{\partial}{\partial N}({\cal A}N^2-{\cal B}) \ne 0, \quad
  {\cal A}\ne 0.
\end{equation}

Then the action \eqref{act_trans_full} becomes (see Appendix \ref{sec:full-U} for more details)
\begin{equation}
\tilde I_{\rm EH}^{\rm unitary} =
\int dt \, d^d x N \sqrt{\gamma} \bigg[ A_4(t,N) \left( K^2 - K^i_{\; j} K^j_{\; i} + (d-1) K L + \frac{d(d-1)}{4} L^2 \right)
- U(t,N,\gamma) \bigg] \; ,
\label{act_trans_unitary}
\end{equation}
up to total derivatives. Here, we have defined
\begin{subequations}
\label{defs_unitary}
\begin{align}
& K_{ij} \equiv \frac{1}{2N} \left( \dot \gamma_{ij} - D_i N_j - D_j N_i \right) \; , \quad
K \equiv \gamma^{ij} K_{ij} \; \\
& L \equiv 
\frac{{\cal A}_N}{\cal A} \left( \frac{\dot N}{N} - \frac{N^i}{N} D_i N \right)  
+ \frac{{\cal A}_t}{{\cal A}N} \; , \\
& U(t,N,\gamma) \equiv - B_4(t,N) \left[ R^{(d)} - (d-1) D^2 \ln {\cal A} - \frac{(d-1)(d-2)}{4} D_i \ln {\cal A} \, D^i \ln {\cal A} \right] \; , \\
& A_4(t,N) \equiv - \frac{M_p^2}{2} \, \frac{N {\cal A}^{d/2}}{\sqrt{{\cal A}N^2-{\cal B}}} \; , \\
& B_4(t,N) \equiv \frac{M_p^2}{2N} {\cal A}^{(d-2)/2} \sqrt{{\cal A}N^2-{\cal B}} \; ,
\end{align}  
\end{subequations}
%%%%
where we have adopted the notation $A_4$ and $B_4$ introduced in the literature~\cite{Gleyzes:2014dya}, $D_i$ and $R^{(d)}$ respectively are the covariant derivative and the Ricci scalar associated with the $d$-dimensional spatial metric $\gamma_{ij}$. We have further defined $D^2 \equiv \gamma^{ij} D_i D_j$, $\dot N \equiv dN/dt$, $\dot \gamma_{ij} \equiv d\gamma_{ij} / dt$, ${\cal A}_t \equiv (\partial {\cal A} / \partial t) \, |_N $, and ${\cal A}_N\equiv (\partial {\cal A} / \partial N) \, |_t$. As one can observe from \eqref{act_trans_unitary}, the transformed action contains $L^2 \supset {\cal A}_N^2 \dot N^2 / ({\cal A}^2 N^2)$ and $KL \supset {\cal A}_N \, \gamma^{ij} \dot \gamma_{ij} \, \dot N / (2 {\cal A} N^2)$, which makes $N$ falsely appear to be a dynamical quantity, provided that ${\cal A}_N \ne 0$. Note that in the general gauge this is equivalent to $ (\partial {\cal A} / \partial X) \, |_\phi  \neq 0$. Nevertheless, this is merely an artifact of the change of variables, as we later show from the full Hamiltonian analysis in the unitary gauge in Sec.~\ref{sec:analysis}, and the number of propagating degrees of freedom is unchanged by the transformation, as one may naively and correctly expect. The unitary-gauge action \eqref{act_trans_unitary} can also be obtained by first fixing the gauge in the original action \eqref{EHaction} and then transforming it in this gauge (see Appendix \ref{sec:full-U} and \ref{sec:unitary}).

In the following section, we perform the Hamiltonian analysis for the transformed action in order to study its structure and to verify the number of physical degrees of freedom. In doing so, we need to introduce the sector associated with $\phi$ in the original action \eqref{EHaction}. As a simple example, we consider the $k$-essence type action, namely,
\begin{equation}
\tilde I_{\rm total} = \tilde I_{\rm EH} + \int d^{d+1}x \sqrt{-\tilde g} \, \tilde P (\phi, \tilde X) \; .
\label{kessence_general}
\end{equation}
After the disformal transformation \eqref{eqn:frametr}, it takes the form
\begin{equation}
\tilde I_{\rm total} = \tilde I_{\rm EH} + \int d^{d+1}x \sqrt{-g} \, {\cal A}^{(d+1)/2} \left( 1 - \frac{\cal B}{\cal A} X \right)^{1/2} \, \tilde P \left( \phi , \frac{X}{{\cal A} - {\cal B} X} \right) \; ,
\end{equation}
where the transformed $\tilde I_{\rm EH}$ is given in \eqref{act_trans_full}.
In the unitary gauge, this %then
becomes
\begin{equation}
\tilde I_{\rm total}^{\rm unitary} = \tilde I_{\rm EH}^{\rm unitary} 
+ \int dt d^dx \, N \sqrt{\gamma} \, A_2(t,N)
\label{kessence_unitary}
\end{equation}
where $\tilde I_{\rm EH}^{\rm unitary}$ is given in \eqref{act_trans_unitary}, and
\begin{equation}
A_2(t,N) \equiv {\cal A}^{(d+1)/2} \left( 1 - \frac{\cal B}{{\cal A} N^2} \right)^{1/2} \, \tilde P \; .
\end{equation}

\section{Hamiltonian analysis for simple case}
\label{sec:analysis}

The transformed action we have derived in Sec.~\ref{sec:frametr} includes the terms quadratic in higher-order derivatives of $\phi$ with respect to time in the general gauge \eqref{act_trans_full} or the terms quadratic in time derivatives of $N$ in the unitary gauge \eqref{act_trans_unitary}, provided ${\cal A}_N \ne 0$. Clearly those terms do not belong to the Horndeski theory or other known theories with the same number of degrees of freedom as that of GR plus one scalar (three in the $4$-dimensional spacetime). Hence, the transformed action appears to have extra degrees of freedom and to become unstable. On the other hand, it is evident that there is no such pathological extra degrees of freedom in the Einstein frame action (\ref{kessence_general}). To investigate whether any extra degrees of freedom really emerge in the transformed system, described by the action \eqref{kessence_unitary}, we shall perform the Hamiltonian analysis in the unitary gauge. 

First we define the canonical momenta conjugate to the variables $N$, $N^i$ and $\gamma_{ij}$ as
\begin{subequations}
\begin{align}
    \pi_N& = \frac{\delta \tilde{I}^{\text{unitary}}_{\text{EH}}}{\delta \dot{N}} = 
               \frac{{\cal A}_N}{{\cal A}}\sqrt{\gamma}A_4(d-1)\left(K+\frac{d}{2}L\right)\ , \\
    \pi_i& = \frac{\delta \tilde{I}^{\text{unitary}}_{\text{EH}}}{\delta \dot{N^i}}=0  \ , \\
    \label{pij}
    \pi^{ij}& =  \frac{\delta \tilde{I}^{\text{unitary}}_{\text{EH}}}{\delta \dot{\gamma_{ij}}}= 
                \sqrt{\gamma}A_4\left(K\gamma^{ij}-K^{ij}+\frac{d-1}{2}\gamma^{ij}L\right)\ ,
\end{align}  
\end{subequations}
where $K_{ij}$, $K$, $L$ and $A_4$ are defined in \eqref{defs_unitary}. The trace of $\pi^{ij}$ is 
\begin{equation}
\pi \equiv \gamma_{ij}\pi^{ij} = \sqrt{\gamma}A_4(d-1)\left(K+\frac{d}{2}L\right) \ , 
\end{equation}
which leads to the relation
\begin{equation}
\pi_N -\frac{{\cal A}_N}{{\cal A}}\pi = 0 \ .
\end{equation}
Using the trace $\pi$, we can rewrite (\ref{pij}) as
\begin{align}
&\pi^{ij} -\frac{1}{d} \, \pi \gamma^{ij} = \sqrt{\gamma}A_4\left(\frac{K}{d}\gamma^{ij}-K^{ij}\right) = {\cal G}^{ij,kl}K_{kl} \ ,
\end{align}
with
\begin{align}
&{\cal G}^{ij,kl} \equiv \sqrt{\gamma}A_4 \left[ \frac{1}{d}\gamma^{ij}\gamma^{kl}- \frac{1}{2}(\gamma^{ik}\gamma^{jl}+\gamma^{il}\gamma^{jk}) \right] \ .   
\end{align}
We cannot invert $\pi^{ij}$ since there is no tensorial product which maps ${\cal G}^{ij,kl}$ to a unit tensorial product. Nevertheless, we can construct the Hamiltonian by the Legendre transformation of the Lagrangian density since the kinetic terms in the Lagrangian have the convex form. Following the usual procedure, the Hamiltonian corresponding to the action \eqref{kessence_unitary} becomes
\begin{align}
H &= \int d^dx\left( \pi^{ij}\dot{\gamma}_{ij} + \pi_N\dot{N} - {\cal L} \right) \nonumber \\
    &= \int d^dx\left( {\cal H}_\bot + N^i{\cal H}_i^N\right)\ ,
\end{align}
where
\begin{align}
    {\cal H}_{\bot}&=  -N\sqrt{\gamma}\left[ \frac{1}{A_4}\left( \frac{\pi^{ij}\pi_{ij}}{\gamma}
    -\frac{1}{d-1}\frac{\pi^2}{\gamma}\right) + \frac{{\cal A}_t}{N{\cal A}}\frac{\pi}{\sqrt{\gamma}} + A_2 -U(t, N, \gamma)\right] \ , \nonumber \\
    {\cal H}_i^N&= {\cal H}_i + \pi_ND_iN \ , \nonumber \\
    {\cal H}_i &\equiv -2\sqrt{\gamma}D_j\left( \frac{\pi^j_{\; i}}{\sqrt{\gamma}}\right) \ , \nonumber
\end{align}
as long as $A_4\neq0$. As we have found above, the primary constraints in this system are
\begin{equation}
\pi_i \approx 0 \ , \quad  \pi_N -\frac{{\cal A}_N}{{\cal A}}\pi \equiv \tilde{\pi}_N \approx 0 \ .
\end{equation}
We define the Poisson brackets as
\begin{align}
 &\{ F,G\}_P \equiv \int d^dx\left( \frac{\delta F}{\delta \Phi^A(x)}\frac{\delta G}{\delta \Pi_A(x)} - \frac{\delta F}{\delta \Pi_A(x)}\frac{\delta G}{\delta \Phi^A(x)}\right ) \ ,
\end{align}
 where
 \begin{align} 
\{ ({\Phi}^A, \Pi_A) \}
 = \Bigl\{ (N, \pi_N), \ (N^i, \pi_i),\ (\gamma_{ij}, \pi^{ij}) \Bigr\} \ .
 \end{align}
The Poisson brackets between the primary constraints yield
\begin{equation}
 \{\pi_i, \pi_j\}_P= 0\,, \quad \{ \pi_i, \tilde{\pi}_N\}_P = 0\,, \quad \{\tilde{\pi}_N, \tilde{\pi}_N\}_P = 0 \ .
\label{eq:btw prim}
\end{equation}
Including the terms corresponding to the primary constraints, we thus redefine the Hamiltonian as
\begin{equation}
 H^\prime = H + \int d^dx (\lambda^i\pi_i + \tilde{\lambda}_N\tilde{\pi}_N),
\end{equation}
where $\lambda^i$ and $\tilde{\lambda}_N$ are Lagrange multipliers.

The consistency conditions of the primary constraints with the time evolution lead us to the secondary constraints as, namely
\begin{align}
\dot{\pi}_i(x) & \approx \{ \pi_i(x), H^\prime\}_P = -{\cal H}_i^N(x)\approx 0 
\end{align}
 and
 \begin{align} 
\dot{\tilde{\pi}}_N(x) &\approx \frac{\partial}{\partial t}\tilde{\pi}_N(x) + \{ \tilde{\pi}_N(x), H^\prime\}_P
\equiv {\cal C} \approx 0 \ , 
 \end{align}
where ${\cal C}$ is given by
\begin{align}
&{\cal C} =  \sqrt{\gamma} D_i\left( N^i\frac{\tilde{\pi}_N}{\sqrt{\gamma}}\right) + \frac{1}{\sqrt{\gamma}}\left[ \left(\frac{N}{A_4}\right)_N + \frac{d{\cal A}_N N}{2{\cal A}A_4}\right]
\left( \pi_{ij}\pi^{ij} -\frac{1}{d-1}\pi^2\right) +  {\cal C}_U[t, N, \gamma, A_{2N}, B_{4N}] \ , \\
&{\cal C}_U =  
\left( \frac{\delta}{\delta N(x)}- \frac{{\cal A}_N}{{\cal A}}\gamma_{ij}\frac{\delta}{\delta \gamma_{ij}(x)}\right)\int d^dyN\sqrt{\gamma}\, \left(A_2-U(t, N, \gamma)\right)\ .
\end{align}

Let us calculate the remaining Poisson brackets among the constraints. Those involving $\pi_i$ are
\begin{align}
    &\{ \pi_i, {\cal H}_j^N \}_P= 0,   \\
    &\{\overline{\pi}[f],\overline{\cal C}[\varphi]\}_P = -\int d^dx \frac{\delta \overline{{\cal C}}[\varphi]}{\delta N^i(x)}f^i = \overline{\tilde{\pi}_N}[f\partial \varphi] \approx 0 \ , \quad \text{for }{}^\forall f^i, {}^\forall \varphi \ ,
\end{align}
where $f\partial\varphi= f^i\partial_i\varphi$ and the barred objects are defined as
\begin{equation}
\overline{\pi}[f] \equiv \int d^dx\pi_if^i \ , \quad
 \overline{\mathcal{C}}[\varphi] \equiv \int d^dx\, \mathcal{C}\, \varphi\, .
\end{equation}
Similarly, we shall make use of the following notation.
\begin{equation}
 \overline{\mathcal{H}}[f] \equiv \int d^dx\, \mathcal{H}_if^i\,, \quad
 \overline{\mathcal{H}^N}[f] \equiv \int d^dx\, \mathcal{H}^N_if^i\, , \quad
 \overline{\tilde{\pi}_N}[\varphi] \equiv \int d^dx\, \tilde{\pi}_N\, \varphi.
\end{equation}
The Poisson brackets involving ${\cal H}^N_i$ are
\begin{subequations}
  \label{HH}
\begin{align}
    &\{ \overline{{\cal H}^N}[f], \overline{{\cal H}^N}[g]\}_P= \overline{{\cal H}^N}[[f,g]]\approx 0\ , \quad \text{for }{}^\forall f^i, {}^\forall g^i \ , \\
    &\{\overline{{\cal H}^N}[f],\overline{\tilde{\pi}_N}[\varphi]\}_P = \overline{\tilde{\pi}_N}[f\partial\varphi] \approx 0   \ , \quad \text{for }{}^\forall f^i, {}^\forall \varphi \ , \\
    &\{\overline{{\cal H}^N}[f], \overline{{\cal C}}[\varphi]\}_P= \overline{{\cal C}}[f\partial\varphi] + \int d^dx \frac{\delta \overline{{\cal C}}[\varphi]}{\delta N^i(x)}[f, N]^i
           = \overline{{\cal C}}[f\partial\varphi] -\overline{\tilde{\pi}_N}\left[[f,N]\partial \varphi\right] \approx 0\ , \quad \text{for }{}^\forall f^i, {}^\forall \varphi \ ,
\end{align}  
\end{subequations}
where $[f, g]^i= f^j\partial_jg^i-g^j\partial_jf^i$ is the Lie bracket of arbitrary $d$-dimensional vectorial quantities $f^i$ and $g^i$. When we calculate (\ref{HH}), we make use of the following formula:
\begin{equation}
\{\overline{{\cal H}}[f], I\}_P =\int d^dx \left\{ \sqrt{\gamma}\frac{\delta I}{\delta \pi_N}D_i\left( f^i\frac{\pi_N}{\sqrt{\gamma}}\right)+ \frac{\delta I}{\delta s}f^iD_is+ \frac{\delta I}{\delta V^i}[f, V]^i\right\} \ ,   
\end{equation}
where $I=I[\pi_N/\sqrt{\gamma}, \gamma_{ij}, \pi^{ij}/\sqrt{\gamma}, 
s, V^i]$ is invariant under the restricted spatial diffeomorphism
\begin{equation}
\label{RSB}
x^i\rightarrow x'^i= x'^i(\vec{x}) \ ,
\end{equation} 
and $s$ and $V^i$ which involve no conjugate momenta transform as a scalar and a vector, respectively, for the restricted spatial diffeomorphism. We note that under this transformation, the quantities $N, \pi_N/\sqrt{\gamma}$ are scalars, $N^i$ is a vector and $\gamma_{ij}, \pi^{ij}/\sqrt{\gamma}$ are rank-$2$ tensors. This formula is a slightly modified version of the one given in Appendix A of \cite{Mukohyama:2015gia}. Lastly, the Poisson bracket of $\tilde{\pi}_N$ and ${\cal C}$ becomes 
\begin{align}
\label{RB8}
   &\{ \tilde{\pi}_N, \overline{{\cal C}}[\varphi]\}_P
      = - \frac{\varphi}{\sqrt{\gamma}}\left\{ \left[ \left(\frac{N}{A_4}\right)_N+\frac{d{\cal A}_NN}{2{\cal A}A_4}\right] _N +\frac{d{\cal A}_N}{2{\cal A}}\left[ \left(\frac{N}{A_4}\right)_N+\frac{d{\cal A}_N N}{2{\cal A}A_4}\right]\right\}
 \left( \pi_{ij}\pi^{ij} -\frac{1}{d-1}\pi^2\right) 
 \nonumber \\ &\qquad\qquad\qquad
 + {\cal C}_U'[t, N, \gamma, A_{2NN}, B_{4NN}] \ , 
\end{align}
where
\begin{align}
 &{\cal C}_U'= - \left( \frac{\delta}{\delta N(x)}-\frac{{\cal A}_N}{{\cal A}}\gamma_{ij}\frac{\delta}{\delta \gamma_{ij}(x)}\right)\int d^{d}y\varphi \, {\cal C}_U \ .   
\end{align}
The Poisson bracket (\ref{RB8}) consists of terms quadratic in $\pi_{ij}$ and the term ${\cal C}_U'$, which does not include any conjugate momenta, and thus Eq.~(\ref{RB8}) would be expressed by nothing but ${\cal C}$ if we could write it by a linear combination of constraints. Equation (\ref{RB8}) includes, however, the second-order partial derivatives of $A_2, A_4$ and $B_4$ with respect to $N$. As a result, we cannot express Eq.~(\ref{RB8}) by a scalar multiple of ${\cal C}$, and it does not vanish weakly in general. Including the terms corresponding to the secondary constraints, we thus redefine the Hamiltonian as
\begin{equation}
\hat{H} = H + \int d^dx \left( \lambda^i \pi_i + \lambda^i_{\cal H} {\cal H}_i^N + \tilde{\lambda}_N \tilde{\pi}_N + \lambda_{\cal C} {\cal C} \right) \; ,
\end{equation}
where $\lambda^i, \lambda^i_{\cal H}, \tilde{\lambda}_N, \lambda_{\cal C}$ are Lagrange multipliers.

The consistency of $\tilde{\pi}_N$ and ${\cal C}$ with the time evolution generated by $\hat{H}$ lead to the following conditions. 
\begin{align}
 \dot{\tilde{\pi}}_N(x) &\approx \frac{\partial}{\partial t}\tilde{\pi}_N(x) + \{ \tilde{\pi}_N(x), H\}_P + \{ \tilde{\pi}_N(x), \overline{{\cal C}}[\lambda_{\cal C}]\}_P  \approx 0 \ , \\
 \dot{{\cal C}}(x) &\approx \frac{\partial}{\partial t}{\cal C}(x) + \{ {\cal C}(x), H\}_P + \{ {\cal C}(x), \overline{\tilde{\pi}_N}[\tilde{\lambda}_N]\}_P+ \{ {\cal C}(x), \overline{{\cal C}}[\lambda_{\cal C}]\}_P  \approx 0 \, ,
\end{align}
which yield no additional constraints but equations to determine the Lagrange multipliers $\tilde{\lambda}_N$ and $\lambda_{\cal C}$. The remaining consistency conditions are
\begin{align}
\dot{\overline{\pi}}[f] & \approx \big\{ \overline{\pi}[f] , \hat{H} \big\}_P \approx - \overline{\cal H}^N [f] + \overline{\tilde \pi_N}[f\partial \lambda_{\cal C}] \approx 0, \\
 \dot{\overline{{\cal H}^N}}[f] & \approx \{ \overline{{\cal H}^N}[f], \hat{H}\}_P \approx 
 \overline{{\cal H}^N}\big[[f,N + \lambda_{\cal H}]\big] + \overline{\tilde \pi_N}\big[ f\partial\lambda_N \big] + \overline{\cal C} \big[f\partial\lambda_{\cal C}\big] \approx 0.
\end{align}
Again, neither of them yields tertiary constraints. Therefore we have $2d$ first-class constraints $\pi_i, {\cal H}^N_i$ and $2$ second-class constraints $\tilde{\pi}_N$ and ${\cal C}$.

Since the dimension of the original phase space ($\gamma_{ij}, \pi^{ij}, N^i, \pi_i, N, \pi_N$) is 
\begin{align}
 \frac{d (d + 1)}{2} \times 2 ~ + ~ d \times 2 ~ + ~ 1 \times 2
 = d^2 + 3 d + 2 \,,  
\end{align}
and there are $2d$ first-class constraints and $2$ second-class constraints, the number of physical degrees freedom of this system is 
\begin{equation}
\# = \frac{1}{2}\left[(d^2+3d+2)-2\times2d-2\right] = \frac{d (d-1)}{2} \ ,
\label{number}
\end{equation}
which is equal to the number of tensor degrees of freedom in $d + 1$ dimension plus that of a scalar field, namely $(d + 1) (d - 2)/2 + 1 = d (d - 1)/2$. With this fact, we conclude that there exist no extra degrees of freedom which may invoke ghost instabilities, although the action includes the terms quadratic in higher-order time derivatives of $\phi$ or the terms quadratic in time derivatives of $N$ in the unitary gauge. This confirms that the number of degrees of freedom in the system with a $k$-essential scalar field coupled to GR does not change under any kind of disformal transformations as long as the transformation is regular and invertible.

\section{General analysis}
\label{sec:general}

So far we have analyzed a concrete example of the theory, transformed from a simple action \eqref{kessence_general} through the transformation \eqref{eqn:frametr}. In this section, we extend our previous result to more general gravitational theories and transformations.

Let us consider a theory of gravity in $d+1$ dimensions described by a set of field components $\{\Phi^A\}$ ($A=1,2,\cdots,\mathcal{N}$), where $\mathcal{N}$ is the total number of field components. For a given theory $\mathcal{N}$ is not unique since the set of field components may include some redundant degrees. For example, $\{\Phi^A\}=\{g_{\mu\nu},\phi\}$ for a scalar tensor theory without any gauge fixing, where $g_{\mu\nu}$ is a metric and $\phi$ is a scalar field, and thus $\mathcal{N}=d(d+1)/2+1$. If we adopt the unitary gauge and the ADM decomposition, on the other hand, the same scalar-tensor theory can be described by the lapse, the shift and the $d$-dimensional spatial metric only, i.e. $\{\Phi^A\}=\{ N, N^i, \gamma_{ij} \}$, and thus $\mathcal{N}=d(d+1)/2$.

In some cases, one might like to introduce auxiliary fields to decrease the order of derivatives in the action. To illustrate this point, suppose that the action of a scalar-tensor theory depends on up to $m$-th order derivatives of the curvature tensor $R_{\alpha\beta\gamma\delta}$ and up to $n$-th order derivatives of a scalar field $\phi$ as
\begin{equation}
 I = \int d^{d+1}x\, L(g_{\mu\nu}, R_{\alpha\beta\gamma\delta}, 
  \nabla_{\mu}R_{\alpha\beta\gamma\delta}, \cdots, 
  \nabla_{(\mu_1}\cdots\nabla_{\mu_m)}R_{\alpha\beta\gamma\delta}, 
  \phi, \nabla_{\mu}\phi, \cdots, 
  \nabla_{(\mu_1}\cdots\nabla_{\mu_n)}\phi), 
  \label{eqn:action-higherderivatives}
\end{equation}
where round brackets denote symmetrization of indices. In this case, if one likes, one can introduce auxiliary fields $\mathcal{R}_{\alpha\beta\gamma\delta}$, $\mathcal{R}_{\mu\alpha\beta\gamma\delta}$, $\cdots$, $\mathcal{R}_{\mu_1\cdots\mu_m\alpha\beta\gamma\delta}$, $\phi_{\mu}$, $\cdots$, $\phi_{\mu_1\mu_2\cdots\mu_n}$ and replace $R_{\alpha\beta\gamma\delta}$ with $\mathcal{R}_{\alpha\beta\gamma\delta}$, $\nabla_{(\mu_1}\mathcal{R}_{\mu_2\cdots\mu_i)\alpha\beta\gamma\delta}$ with $\mathcal{R}_{\mu_1\cdots\mu_i\alpha\beta\gamma\delta}$ ($i=1,\cdots,n$) and $\nabla_{(\mu_1}\phi_{\mu_2\cdots\mu_j)}$ with $\phi_{\mu_1\cdots\mu_j}$ ($j=1,\cdots,n$), provided that corresponding Lagrange multipliers are properly introduced. The new action, which is equivalent to (\ref{eqn:action-higherderivatives}), is then
\begin{eqnarray}
 I' & = & \int d^{d+1}x\, 
  \left[
   L(g_{\mu\nu}, \mathcal{R}_{\alpha\beta\gamma\delta}, 
  \mathcal{R}_{\mu\alpha\beta\gamma\delta}, \cdots, 
  \mathcal{R}_{\mu_1\cdots\mu_m\alpha\beta\gamma\delta}, 
  \phi, \phi_{\mu}, \cdots, 
  \phi_{\mu_1\cdots\mu_n})
  + \Lambda^{\alpha\beta\gamma\delta}
  (\mathcal{R}_{\alpha\beta\gamma\delta}-R_{\alpha\beta\gamma\delta})
  \right.
  \nonumber\\
 & & 
  \left.
  + \Lambda^{\mu\alpha\beta\gamma\delta}
  (\mathcal{R}_{\mu\alpha\beta\gamma\delta}-\nabla_{\mu}\mathcal{R}_{\alpha\beta\gamma\delta})
  + \cdots
  + \Lambda^{\mu_1\cdots\mu_m\alpha\beta\gamma\delta}
  (\mathcal{R}_{\mu_1\cdots\mu_m\alpha\beta\gamma\delta}
  -\nabla_{(\mu_m}\mathcal{R}_{\mu_1\cdots\mu_{m-1})\alpha\beta\gamma\delta})
  \right.
  \nonumber\\
 & & 
  \left.
  + \lambda^{\mu}(\phi_{\mu}-\nabla_{\mu}\phi)
  + \cdots 
  + \lambda^{\mu_1\cdots\mu_n}
  (\phi_{\mu_1\cdots\mu_n}-\nabla_{(\mu_m}\phi_{\mu_1\cdots\mu_{m-1})})
  \right],
\end{eqnarray}
where $\Lambda^{\alpha\beta\gamma\delta}$, $\Lambda^{\mu_1\cdots\mu_i\alpha\beta\gamma\delta}$ ($i=1,\cdots,m$) and $\lambda^{\mu_1\cdots\mu_j}$ ($j=1,\cdots,n$) are Lagrange multipliers. This action is linear in the curvature $R_{\alpha\beta\gamma\delta}$. Since the curvature is linear in the second derivatives of the metric components, one can perform an integration by parts to remove the terms depending on the second derivatives. Concretely, the action after the integration by parts is
\begin{eqnarray}
 I' & = & \int d^{d+1}x\, 
  \left[
   L(g_{\mu\nu}, \mathcal{R}_{\alpha\beta\gamma\delta}, 
  \mathcal{R}_{\mu\alpha\beta\gamma\delta}, \cdots, 
  \mathcal{R}_{\mu_1\cdots\mu_m\alpha\beta\gamma\delta}, 
  \phi, \phi_{\mu}, \cdots, 
  \phi_{\mu_1\cdots\mu_n})
  \right.
  \nonumber\\
 & & 
  \left.
  + \Gamma^{\alpha}_{\ \beta\delta}\partial_{\gamma}
  (g_{\alpha\eta}\Lambda^{\eta\beta\gamma\delta})
  - \Gamma^{\alpha}_{\ \beta\gamma}\partial_{\delta}
  (g_{\alpha\eta}\Lambda^{\eta\beta\gamma\delta})
  + \Lambda^{\alpha\beta\gamma\delta}
  (\mathcal{R}_{\alpha\beta\gamma\delta}
  -g_{\alpha\eta}\Gamma^{\eta}_{\ \epsilon\gamma}\Gamma^{\epsilon}_{\ \beta\delta}
  +g_{\alpha\eta}\Gamma^{\eta}_{\ \epsilon\delta}\Gamma^{\epsilon}_{\ \beta\gamma})
  \right.
  \nonumber\\
 & & 
  \left.
  + \Lambda^{\mu\alpha\beta\gamma\delta}
  (\mathcal{R}_{\mu\alpha\beta\gamma\delta}-\nabla_{\mu}\mathcal{R}_{\alpha\beta\gamma\delta})
  + \cdots
  + \Lambda^{\mu_1\cdots\mu_m\alpha\beta\gamma\delta}
  (\mathcal{R}_{\mu_1\cdots\mu_m\alpha\beta\gamma\delta}
  -\nabla_{(\mu_m}\mathcal{R}_{\mu_1\cdots\mu_{m-1})\alpha\beta\gamma\delta})
  \right.
  \nonumber\\
 & & 
  \left.
  + \lambda^{\mu}(\phi_{\mu}-\nabla_{\mu}\phi)
  + \cdots 
  + \lambda^{\mu_1\cdots\mu_n}
  (\phi_{\mu_1\cdots\mu_n}-\nabla_{(\mu_m}\phi_{\mu_1\cdots\mu_{m-1})})
  \right], \label{eqn:modified-action}
\end{eqnarray}
where $\Gamma^{\alpha}_{\ \beta\delta}$ is the Christoffel symbol for the metric $g_{\mu\nu}$. Therefore, the action after the integration by parts depends on $\{g_{\mu\nu}, \mathcal{R}_{\alpha\beta\gamma\delta}$, $\mathcal{R}_{\mu\alpha\beta\gamma\delta}$, $\cdots$, $\mathcal{R}_{\mu_1\cdots\mu_m\alpha\beta\gamma\delta}$, $\phi$, $\phi_{\mu}$, $\cdots$, $\phi_{\mu_1\cdots\mu_n}$, $\Lambda^{\alpha\beta\gamma\delta}$, $\Lambda^{\mu\alpha\beta\gamma\delta}$, $\cdots$, $\Lambda^{\mu_1\cdots\mu_m\alpha\beta\gamma\delta}$, $\lambda^{\mu}$, $\cdots$, $\lambda^{\mu_1\cdots\mu_n}, \}$ and first derivatives of some of them only. Moreover, the action is at most second order in the first derivatives of fields. It is easy to extend this procedure to multiple scalars, vectors, antisymmetric tensor fields, and so on. If the original action is not (manifestly) diffeomorphism invariant then it may be more convenient (although not mandatory) to covariantize the action by adding St\"uckelberg fields to the system before introducing auxiliary fields and Lagrange multipliers. One may also fix some gauge freedom after rendering the action of the form (\ref{eqn:modified-action}), if it is convenient.

Throughout this section, we suppose that the action of the theory under consideration can be cast into the form
\begin{equation}
 I = \int 
 d^{d+1}x
 \left[\frac{1}{2}\mathcal{K}_{AB}\dot{\Phi}^A\dot{\Phi}^B + M_A\dot{\Phi}^A - V \right], \label{eqn:general-action}
\end{equation}
where $\mathcal{K}_{AB}$ ($=\mathcal{K}_{BA}$), $M_A$ and $V$ are functions of the time $t$, $\Phi^C$ ($C=1,2,\cdots, {\cal N}$) and their spatial derivatives. We might need to introduce auxiliary fields to cast the action into this form, as illustrated in (\ref{eqn:modified-action}). Simple examples of this procedure are depicted in Appendix \ref{app:example}. Hereafter, we assume that the set of variables $\{\Phi^A\}$ includes such auxiliary fields (if necessary) and the action is already cast into the form (\ref{eqn:general-action}).

We consider a transformation of the form,
\begin{equation}
 \tilde{\Phi}^A = F^A(\Phi,t), \quad (A=1,2,\cdots, {\cal N} 
 ),\label{eqn:general-transformation}
\end{equation}
with ${\cal N}$ being the total number of field components in the system and suppose that $\det F^A_B\ne 0$, $\infty$, where $F^A_B=\partial F^A/\partial\Phi^B$. This type of theory \eqref{eqn:general-action} and transformation \eqref{eqn:general-transformation} clearly includes the case we have studied in Secs.~\ref{sec:frametr} and \ref{sec:analysis}, and in the previous example (\ref{eqn:unitary-gauge-transformation}), $\{\Phi^A\}=\{N,N^i,\gamma_{ij}\}$ and $\{\tilde{\Phi}^A\}=\{\tilde{N},\tilde{N}^i,\tilde{\gamma}_{ij}\}$. The transformation (\ref{eqn:general-transformation}) is more general than the previous example but still does not involve derivatives of the field components $\{\Phi^A\}$. 

If the transformation under consideration involves derivatives of the field components $\{\Phi^A\}$ then the number of physical degrees of freedom may change in general. In some special cases, however, such a transformation can be cast into the form (\ref{eqn:general-transformation}) by introducing further auxiliary fields and Lagrange multipliers as we did in the derivation of (\ref{eqn:modified-action}). Finding the necessary and sufficient condition for this to be possible is an interesting problem but is beyond the scope of the present paper. Here we simply conjecture that this should be possible if a derivative-dependent transformation is regular and invertible in the sense that it can be inverted to express $\{\Phi^A\}$ in terms of $\{\tilde{\Phi}^A\}$ and their derivatives. (See Appendix~\ref{appsub:regular} for a simple example of this kind. As shown in (\ref{eqn:inverse-transformation}), the frame transformation (\ref{eqn:frametr}) also falls into this type if ${\cal A}({\cal A}-{\cal B}X)({\cal A}-{\cal A}_XX+{\cal B}_XX^2)\neq 0$.) After introduction of the further auxiliary fields and Lagrange multipliers, the action is still of the form (\ref{eqn:general-action}) with a larger $\mathcal{N}$. Another example is the case in which the transformation becomes derivative-independent one in a special gauge. In this case one can simply perform a canonical transformation, corresponding to the gauge transformation from a general gauge to the special gauge, to cast the transformation under consideration into the form (\ref{eqn:general-transformation}).

Since we assume that $\det F^A_B\ne 0$, $\infty$, the transformation (\ref{eqn:general-transformation}) can be inverted as 
\begin{equation}
 \Phi^A = G^A(\tilde{\Phi},t), \quad (A=1,2,\cdots, {\cal N}
 ), \label{eqn:Phi=G}
\end{equation}
with $\det G^A_B\ne 0$, $\infty$, where $G^A_B=\partial G^A/\partial\tilde{\Phi}^B$. Since
\begin{eqnarray}
d\Phi^A & = & G^A_CF^C_Bd\Phi^B + (G^A_B\partial_tF^B+\partial_tG^A)dt, \nonumber\\
d\tilde{\Phi}^A & = & F^A_CG^C_Bd\tilde{\Phi}^B + (F^A_B\partial_tG^B+\partial_tF^A)dt, 
\end{eqnarray}
we have the following identities.
\begin{equation}
 G^A_CF^C_B = F^A_CG^C_B = \delta^A_B, \quad
  G^A_B\partial_tF^B+\partial_tG^A = 
  F^A_B\partial_tG^B+\partial_tF^A = 0.
  \label{eqn:identities1}
\end{equation}
In addition, by taking the variation of both sides of $G^A_CF^C_B=\delta^A_B$, one obtains
\begin{equation}
 (G^A_{CD}F^C_B+G^A_CF^C_{BE}G^E_D)d\tilde{\Phi}^D + (F^C_B\partial_tG^A_C+G^A_CF^C_{BE}\partial_tG^E+G^A_C\partial_tF^C_B)dt = 0, 
\end{equation}
which yields the identities
\begin{equation}
 G^A_{CD}F^C_B+G^A_CF^C_{BE}G^E_D = 0, \quad
  F^C_B\partial_tG^A_C+G^A_CF^C_{BE}\partial_tG^E+G^A_C\partial_tF^C_B = 0,
  \label{eqn:identities2}
\end{equation}
where $G^A_{CD}=\partial^2G^A/\partial\tilde{\Phi}^C\partial\tilde{\Phi}^D$ and
$F^C_{BE}=\partial^2F^C/\partial\Phi^B\partial\Phi^E$.

After the transformation $\{\Phi^A\}\to\{\tilde{\Phi}^A\}$, the action (\ref{eqn:general-action}) becomes
\begin{equation}
 I = \int
 d^{d+1}x \left[\frac{1}{2}\tilde{\mathcal{K}}_{AB}\dot{\tilde{\Phi}}^A\dot{\tilde{\Phi}}^B + \tilde{M}_A\dot{\tilde{\Phi}}^A - \tilde{V} \right],
\end{equation}
where
\begin{eqnarray}
 \tilde{\mathcal{K}}_{AB} & = & \mathcal{K}_{CD}G^C_AG^D_B, \nonumber\\
 \tilde{M}_A & = & M_BG^B_A + \mathcal{K}_{CB}G^C_A\partial_tG^B, \nonumber\\
 \tilde{V} & = & V-M_A\partial_tG^A-\frac{1}{2}\mathcal{K}_{AB}\partial_tG^A\partial_tG^B.
\end{eqnarray}
It is immediate to see that
\begin{equation}
 \mathrm{rank}\,\mathcal{K}_{AB} =  \mathrm{rank}\,\tilde{\mathcal{K}}_{AB}\,,
  \label{eqn:rankK=ranktildeK}
\end{equation}
since $\mathcal{K}_{AB}v_{\alpha}^B=0$ ($\alpha=1,2,\cdots, {\cal N}-\mathrm{rank}\,\mathcal{K}_{AB}$) implies $\tilde{\mathcal{K}}_{AB}\tilde{v}_{\alpha}^B=0$, where $\tilde{v}_{\alpha}^A=F^A_Bv_{\alpha}^B$, and $\tilde{\mathcal{K}}_{AB}\tilde{w}_{\nu}^B=0$ ($\nu=1,2,\cdots, {\cal N}-\mathrm{rank}\,\tilde{\mathcal{K}}_{AB}$) implies $\mathcal{K}_{AB}w_{\alpha}^B=0$, where $w_{\nu}^A=G^A_B\tilde{w}_{\nu}^B$. Here $\{v_{\alpha}|\alpha=1,2,\cdots, {\cal N}-\mathrm{rank}\,\mathcal{K}_{AB}\}$ is the set of independent eigenvectors of $\mathcal{K}_{AB}$ with zero eigenvalues, and $\{w_{\nu}|\nu=1,2,\cdots, {\cal N}-\mathrm{rank}\,\tilde{\mathcal{K}}_{AB}\}$ is the set of independent eigenvectors of $\tilde{\mathcal{K}}_{AB}$ with zero eigenvalues.

One can define the canonical momenta $\Pi_A$ conjugate to $\Phi^A$ in the standard way as 
\begin{equation}
 \Pi_A = \frac{\delta I}{\delta \dot{\Phi}^A} = \mathcal{K}_{AB}\dot{\Phi}^B + M_A. 
\end{equation}
Hence, if $r=\mathrm{rank}\,\mathcal{K}_{AB}$ is less than ${\cal N}$, then there are ${\cal N}-r$ primary constraints, which are given by 
\begin{equation}
 \mathcal{C}_{\alpha} \equiv v_{\alpha}^A(\Pi_A-M_A) = 0, \quad (\alpha=1,2,\cdots ,
 {\cal N}
 -r)\,. 
\end{equation}
Similarly, the canonical momenta $\tilde{\Pi}_A$ conjugate to $\tilde{\Phi}^A$ for the system after the transformation (\ref{eqn:general-transformation}) are
\begin{equation}
 \tilde{\Pi}_A = \frac{\delta I}{\delta \dot{\tilde{\Phi}}^A} = \tilde{\mathcal{K}}_{AB}\dot{\tilde{\Phi}}^B + \tilde{M}_A = \Pi_BG^B_A, \label{enq:tildePi}
\end{equation}
where we have used the identities (\ref{eqn:identities1}) to show the last equality.
Clearly there are again ${\cal N}-r$ primary constraints, as in the new frame we have
\begin{equation}
 \tilde{\mathcal{C}}_{\alpha} \equiv \tilde{v}_{\alpha}^A(\tilde{\Pi}_A-\tilde{M}_A) = 0, \quad (\alpha=1,2,\cdots ,
 {\cal N} -r)\,,  \label{eqn:tildeCalpha}
\end{equation}
where $\tilde{v}_{\alpha}^A=F^A_Bv_{\alpha}^B$. Because of (\ref{eqn:rankK=ranktildeK}), it is obvious that (\ref{eqn:tildeCalpha}) exhausts all independent primary constraints for the system after the transformation (\ref{eqn:general-transformation}). Moreover, it is straightforward to show that
\begin{equation}
 \mathcal{C}_{\alpha} = \tilde{\mathcal{C}}_{\alpha}. 
\end{equation}
On the other hand, the Hamiltonians for the system before and after the transformation \eqref{eqn:general-transformation}, respectively denoted by $H$ and $\tilde H$, are formally
\begin{equation}
 H = \int 
 d^dx
 \left[\Pi_A\dot{\Phi}^A - \frac{1}{2}\mathcal{K}_{AB}\dot{\Phi}^A\dot{\Phi}^B - M_A\dot{\Phi}^A + V\right],
\end{equation}
and
\begin{equation}
 \tilde{H} = \int 
 d^dx
 \left[\tilde{\Pi}_A\dot{\tilde{\Phi}}^A - \frac{1}{2}\tilde{\mathcal{K}}_{AB}\dot{\tilde{\Phi}}^A\dot{\tilde{\Phi}}^B - \tilde{M}_A\dot{\tilde{\Phi}}^A + \tilde{V} \right], 
\end{equation}
and are related to each other as
\begin{equation}
 \tilde{H} = H - \int d^dx \,\Pi_A\partial_tG^A, \label{eqn:tildeH-H}
\end{equation}
where we have used the identities (\ref{eqn:identities1}).

The Poisson bracket is defined through
\begin{equation}
 \left\{ \Phi^A(\vec{x}), \Phi^B(\vec{y})\right\}_P = 0, \quad
 \left\{ \Phi^A(\vec{x}), \Pi_B(\vec{y})\right\}_P = \delta^A_B\delta^3(\vec{x}-\vec{y}), \quad
 \left\{ \Pi_A(\vec{x}), \Pi_B(\vec{y})\right\}_P = 0. 
\end{equation}
From these, it follows that
\begin{equation}
 \left\{ \tilde{\Phi}^A(\vec{x}), \tilde{\Phi}^B(\vec{y})\right\}_P = 0, \quad
 \left\{ \tilde{\Phi}^A(\vec{x}), \tilde{\Pi}_B(\vec{y})\right\}_P =
 \delta^A_B 
 \delta^3(\vec{x}-\vec{y}), \quad
 \left\{ \tilde{\Pi}_A(\vec{x}), \tilde{\Pi}_B(\vec{y})\right\}_P = 0, 
\end{equation}
meaning that the definition of the Poisson bracket does not change by the transformation (\ref{eqn:general-transformation}).

If one considers the time evolution of a function, let us say
\begin{equation}
 \mathcal{O}(\Phi,\Pi,t) =  \tilde{\mathcal{O}}(\tilde{\Phi},\tilde{\Pi},t),
  \label{eqn:O=tildeO}
\end{equation}
it is easy to show that
\begin{equation}
 \left\{ \tilde{\mathcal{O}}, \tilde{H} \right\}_P
  = \left\{ \mathcal{O}, H \right\}_P
  - \left(\frac{\partial\mathcal{O}}{\partial\Phi^A}\right)_{\Pi,t}\partial_tG^A
  + \left(\frac{\partial\mathcal{O}}{\partial\Pi_A}\right)_{\Phi,t}\Pi_BF^C_A\partial_tG^B_C. \label{eqn:OH}
\end{equation}
By taking variation of both sides of (\ref{eqn:O=tildeO}), one obtains the following identities:
\begin{eqnarray}
 \left(\frac{\partial\tilde{\mathcal{O}}}{\partial\tilde{\Phi}^A}\right)_{\tilde{\Pi},t} & = &  \left(\frac{\partial\mathcal{O}}{\partial\Phi^B}\right)_{\Pi,t}G^B_A
  + \left(\frac{\partial\mathcal{O}}{\partial\Pi_B}\right)_{\Phi,t}\Pi_C G^C_E F^E_{BD}G^D_A,\nonumber\\
 \left(\frac{\partial\tilde{\mathcal{O}}}{\partial\tilde{\Pi}_A}\right)_{\tilde{\Phi},t} & = &  \left(\frac{\partial\mathcal{O}}{\partial\Pi_B}\right)_{\Phi,t}F^A_B, \nonumber\\
 \left(\frac{\partial\tilde{\mathcal{O}}}{\partial t}\right)_{\tilde{\Phi},\tilde{\Pi}} & = &  \left(\frac{\partial\mathcal{O}}{\partial t}\right)_{\Phi,\Pi} + \left(\frac{\partial\mathcal{O}}{\partial\Phi^A}\right)_{\Pi,t}\partial_tG^A + \left(\frac{\partial\mathcal{O}}{\partial\Pi_A}\right)_{\Phi,t}\Pi_DG^D_B(\partial_tF^B_A+F^B_{AC}\partial_tG^C). \nonumber\\
\end{eqnarray}
The last identity, combined (\ref{eqn:OH}) with (\ref{eqn:identities2}), leads to 
\begin{equation}
 \left\{ \tilde{\mathcal{O}}, \tilde{H}\right\}_P + \left(\frac{\partial\tilde{\mathcal{O}}}{\partial t}\right)_{\tilde{\Phi},\tilde{\Pi}}
  =  \left\{ \mathcal{O}, H\right\}_P + \left(\frac{\partial\mathcal{O}}{\partial t}\right)_{\Phi,\Pi}.
\end{equation}
Hence, the time evolution of the system after the transformation (\ref{eqn:general-transformation}) is exactly the same as that before the transformation, provided that $\tilde{H}$ and $H$, respectively, are used as the Hamiltonian of each system.

In summary, the set of all primary constraints, the definition of the Poisson bracket and the time evolution after the transformation (\ref{eqn:general-transformation}) are the same as those before the transformation. The essential reason for this is that the transformation between $\{\Phi^A,\Pi_B\}$ and $\{\tilde{\Phi}^A,\tilde{\Pi}_B\}$ is a canonical transformation. Indeed, in terms of the generating functional 
\begin{equation}
 \mathcal{G}[\Pi,\tilde{\Phi};t] = -\int d^dx\, \Pi_A\,G^A(\tilde{\Phi},t), 
\end{equation}
the transformation (\ref{eqn:Phi=G}) and (\ref{enq:tildePi}) can be cast into the form of the standard canonical transformation as
\begin{equation}
 \Phi^A = -\frac{\delta\mathcal{G}}{\delta\Pi_A}, \quad
  \tilde{\Pi}_A = -\frac{\delta\mathcal{G}}{\delta\tilde{\Phi}^A}, 
\end{equation}
The standard formula for the transformation of Hamiltonians is
\begin{equation}
 \tilde{H} = H + \frac{\partial\mathcal{G}}{\partial t},
\end{equation}
which agrees with (\ref{eqn:tildeH-H}). This means that all the secondary constraints and the constraint algebra are also the same since they are defined through the primary constraints, the Poisson bracket and the time evolution. We thus conclude that the number of degrees of freedom does not change by a generic transformation of the type \eqref{eqn:general-transformation}, provided that $\det F^A_B\ne 0$, $\infty$.

\section{Summary and discussion}
\label{sec:summary}

We have studied metric transformations of the form (\ref{eqn:frametr}), which depend on a scalar field and its first derivatives, for a scalar-tensor gravitational theory with a metric $g_{\mu\nu}$ and a scalar $\phi$. We have explicitly seen that, unless the conformal factor $\mathcal{A}$ is independent of $X=-g^{\mu\nu}\partial_{\mu}\phi\partial_{\nu}\phi$, the equations of motion in the theory after transformation generically involve higher-order derivatives of the scalar field even if we start with a theory whose equations of motion are second-order. One might thus mistakenly conclude that the theory after such transformation would have more physical degrees of freedom than the original one. However, the two theories should be equivalent if the transformation is regular and invertible. In order to reconcile the apparent contradiction, we have applied the standard method of Hamiltonian analysis for a constrained system. In this paper, we have proven that frame transformations of the form (\ref{eqn:frametr}) never change the number of physical degrees of freedom in the unitary gauge where $\phi = t$, provided that the transformation is regular and invertible and that $\partial_\mu \phi$ is non-zero, regular and timelike everywhere in spacetime. In addition, we have checked that the metric transformation and the gauge fixing to the unitary gauge in transforming the action \eqref{EHaction} are commutative, namely that one can fix the unitary gauge before or after the transformation (see Appendix \ref{sec:detailed}). Given the fact that higher order time derivatives of the scalar field disappear in the unitary gauge $\phi = t$ and that the transformations which originally include derivatives of the scalar field reduce to simple point transformations (without any derivatives), taking the unitary gauge, if it exists, enables us to perform the analysis with ease in a rather concise manner. On the other hand, performing a full Hamiltonian analysis is already a rather involved task and hence it will be interesting to seek a more simple and hopefully elegant way instead of it, if any.

Afterwards, we have extended the discussion to more general gravitational theories and transformations in any gauge. We found that the set of all primary constraints, the definition of the Poisson bracket and the time evolution after the transformation (\ref{eqn:general-transformation}) are the same as those before the transformation. This means that all the secondary constraints and the constraint algebra are also the same. Therefore, we conclude that the number of degrees of freedom does not change by a generic transformation, provided that the transformation is regular and invertible.

Here let us make some comments on a singular transformation to which our analysis does not apply, since it violates regularity and invertibility of the transformation. If ${\rm det} \, F^A_{\ B} = 0$ for the transformation of the type \eqref{eqn:general-transformation}, the transformation becomes singular, which can happen, for example, for the disformal transformation (\ref{eqn:frametr}) when there is a particular relation between ${\cal A}$ and ${\cal B}$ violating Eq.~\eqref{eqn:invertibility-condition} \cite{Deruelle:2014eha}\,:
\begin{align}
 {\cal B} (\phi \,, X) = \frac{1}{X} {\cal A} (\phi \,, X) - f (\phi) \; ,
 \label{mimetic}
\end{align}
where $f (\phi)$ is an arbitrary positive function of the scalar field alone. Since the transformation is not invertible in this case ($\tilde{N}^2 = f(t)$ in the unitary gauge), the number of physical degrees of freedom may change by the transformation, which will be the origin of mimetic degree of freedom \cite{Chamseddine:2013kea,Chamseddine:2014vna} (see also \cite{Deruelle:2014eha,nojiri2014mimetic,Arroja:2015wpa}). While we have not investigated this special case in this paper, there are related interesting issues. For example, due to the singular nature of the transformation, the number of degrees of freedom may decrease instead of increase while the latter case is mostly studied so far in the literature. Moreover when the transformation between two theories is singular, whether one of the theories can be a consistent truncation of the other is also an interesting question. In particular for the case of the original theory of Chamseddine and Mukhanov \cite{Chamseddine:2013kea,Chamseddine:2014vna}, the Hamiltonian analysis yields that the physical degrees of freedom in fact increase in number \cite{chaichian2014mimetic,malaeb2015hamiltonian}. However, in the original mimetic gravity the scalar field is non-dynamical, i.e. there is no kinetic term, and the singular transformation leads to a new conformal gauge degree of freedom \cite{chaichian2014mimetic}, a.k.a. the mimetic degree of freedom. Contrariwise, if the scalar field is originally dynamical, it is not clear whether a new degree of freedom appears or not. In any case, it is worth performing a Hamiltonian analysis in a general way as done recently in~\cite{deffayet2015counting} and study the limit of mimetic gravity. We hope to come back to these issues in the near future.

On the other hand, in another approach in the literature \cite{Zumalacarregui:2013pma,bettoni2015kinetic,motohashi2015dis}, it has been shown that despite the appearance of higher-order derivatives in the action one may rewrite the equations of motion, after some manipulations, only in the terms up to second-order derivatives. This fact is usually referred as the existence of implicit or "hidden" constraints. However, such approach is not fully satisfactory from an action point of view and one may be eager to reveal the "hidden" constraint with the help of a Lagrange multiplier \cite{chen2013higher}. In that sense, as shown in Sec.~\ref{sec:general} and inspired in the alternative formulation of mimetic gravity \cite{chaichian2014mimetic,golovnev2014mimetic,Arroja:2015wpa}, one might expect that by treating the seemingly problematic $X$ dependence as a new field, say $\chi$, introduced by a Lagrange multiplier $\lambda$, namely 
\begin{align}
I(g,\phi,\partial_\mu\phi,X,\partial_\mu X) = 
I(g,\phi,\partial_\mu\phi,\chi,\partial_\mu\chi)+\int d^{d+1}x~\lambda(\chi-X),
\end{align}
one is able to get rid of higher-order derivatives at the level of action and simplify the Hamiltonian analysis. The mimetic limit is achieved when $\chi$ becomes a conformal gauge degree of freedom and can be set to unity, thus recovering the results in the literature \cite{chaichian2014mimetic,golovnev2014mimetic,Arroja:2015wpa}, where one has the original action plus a constraint given by~\footnote{One can always redefine the field in such a way that the constraint always takes the same form as in \eqref{eq:mimeticconstraint}.}
\begin{align}\label{eq:mimeticconstraint}
 X = - g^{\mu\nu}\partial_\mu\phi\partial_\nu\phi = 1 \,.
\end{align}
One must bear in mind that this is just speculation, and a careful analysis should be derived in an upcoming work. Nevertheless, some insights are at hand from our Hamiltonian formulation in the unitary gauge. In an $X$-dependent metric transformation the ``hidden" constraint is already revealed at the Hamiltonian level by an unambiguous systematic procedure. In other words, the undesired canonical momentum from the higher-order derivatives, i.e. $\pi_N$, is related to an already existing momentum, in our case $\pi^{ij}$, which gives rise to two additional second class constraints, namely
\begin{align}
 \tilde{\pi}_N=\pi_N- \frac{{\cal A}_N}{{\cal A}}\gamma_{ij} \pi^{ij}\approx 0
\end{align}
and its constancy in time, which eliminates the extra degree of freedom, or in other words one can use these constraints to completely rewrite the problematic higher-order time derivative terms in terms of second-order quantities. This nontrivial constraint, $\tilde{\pi}_N \approx 0$, and the associated secondary one would be the unitary-gauge counterpart of the implicit constraints we would look for in the analysis without fixing the gauge.

From the Hamiltonian analysis in the present paper, it is obvious that there exist more general scalar-tensor theories with the same number of physical degrees of freedom as that in the theories with GR plus one scalar mode. For example, if the action in the unitary gauge is of the form
\begin{equation}
 I = \int 
 d^{d+1}x \,
 N\sqrt{\gamma} \, {\cal L}(K+\alpha \,
 \partial_\perp N, K^{\mathrm{T}}_{ij}, N, \gamma_{ij}, D_i) \,,
\end{equation}
where $\partial_\perp N \equiv \frac{\dot N}{N} - \frac{N^i}{N} \partial_i N$, $\alpha$ is a $d$-dimensional scalar made of $N$, $\gamma_{ij}$ and $D_i$, and 
\begin{equation}
K^{\mathrm{T}}_{ij} = K_{ij}- 
\frac{1}{d} \,K\gamma_{ij}, 
\end{equation}
then there is a primary constraint of the form
\begin{equation}
 \pi_N - 
 \frac{2}{d} \,\alpha \, \gamma_{ij} \pi^{ij} 
 = 0.
\end{equation}
Hence, it is expected that a theory of this type generically has the same number of degrees of freedom as that of general relativity plus one more degree of freedom.

%%%%%%%%%%%%%%%%%%%%%%%%%%%%%%%%%%%%%%%%%%%%%%%%%%%%%%%%%%%%%%%%%%%%%
\acknowledgments
A.N. and G.D. would like to thank Misao Sasaki for many fruitful discussions.
S.M.'s work was supported in part by Grant-in-Aid for Scientific Research 24540256. 
A.N. is grateful to the Yukawa Institute for Theoretical Physics
 at Kyoto University for warm hospitality
 where part of this work was done. 
The work of A.N. was supported in part by the JSPS Research
 Fellowship for Young Scientists No. $263409$. 
Y.W.'s work was supported by the Program for Leading Graduate Schools, MEXT, Japan. 
This work was supported in part by the WPI Initiative, MEXT Japan. 

%%%%%%%%%%%%%%%%%%%%%%%%%%%%%%%%%%%%%%%%%%%%%%%%%%%%%%%%
%%%%%%%%%%%%%%%%%%%%%%%%%%%%%%%%%%%%%%%%%%%%%%%%%%%%%%%%
\appendixpage
\appendix
\section{Detailed calculation of transformation}
\label{sec:detailed}

\subsection{Derivation of transformed action without gauge fixing}
\label{sec:full}

In this appendix subsection, we show some details of the calculation in disformally transforming the Einstein-Hilbert action to obtain \eqref{act_trans_full}. For a general transformation from the metric $\tilde g_{\mu\nu}$ to $g_{\mu\nu}$, it is convenient to know the relation
\begin{equation}
\tilde R_{\mu\nu} - R_{\mu\nu} = \tilde \nabla_\rho \delta \Gamma^\rho_{\mu\nu} + \delta \Gamma^\rho_{\mu\sigma} \delta \Gamma^{\sigma}_{\rho\nu}
- \tilde \nabla_\mu \tilde\nabla_\nu \ln \frac{\sqrt{-\tilde g}}{\sqrt{-\tilde g}} - \delta\Gamma^\rho_{\mu\nu} \nabla_\rho \ln \frac{\sqrt{-\tilde g}}{\sqrt{-g}} \; ,
\end{equation}
where
\begin{equation}
\delta \Gamma^\rho_{\mu\nu} \equiv \tilde \Gamma^\rho_{\mu\nu} - \Gamma^\rho_{\mu\nu} \; ,
\end{equation}
which is a tensorial quantity. Hence the corresponding part of the action reduces to \cite{Zumalacarregui:2013pma}
\begin{equation}
\int d^{d+1}x \sqrt{-\tilde g} \, \tilde g^{\mu\nu} \left( \tilde R_{\mu\nu} - R_{\mu\nu} \right) = 
\int d^{d+1}x \sqrt{-g} \frac{\sqrt{-\tilde g}}{\sqrt{-g}} \, \tilde g^{\mu\nu} \left( \delta \Gamma^\rho_{\mu\sigma} \delta\Gamma^\sigma_{\rho\nu} - \delta \Gamma^\rho_{\mu\nu} \delta \Gamma^\sigma_{\rho\sigma} \right) \; ,
\label{act_trans_general}
\end{equation}
up to total derivatives. This expression is particularly convenient, due to its compactness and to the fact that no derivatives on the Christoffel symbols are involved.

We now consider the conformal and disformal transformations of the metric. Let us write the transformation of $\tilde g_{\mu\nu}$, \eqref{eqn:frametr}, and that of $\tilde g^{\mu\nu}$, \eqref{trans_inverse}, as 
\begin{align}
\tilde g_{\mu\nu} = {\cal A} \left( g_{\mu\nu} + {\cal F} \, \phi_\mu \phi_\nu \right) \; , \quad
\tilde g^{\mu\nu} = \frac{1}{{\cal A}} \left( g^{\mu\nu} - {\cal G} \, \phi^\mu \phi^\nu \right) \; ,
\end{align}
where
\begin{equation}
{\cal F} \equiv \frac{{\cal B}}{{\cal A}} \; , \quad
{\cal G} \equiv \frac{{\cal B}}{{\cal A}-{\cal B}X} \; , \quad 
\phi_\mu \equiv \nabla_\mu \phi \; , \quad
\phi^\mu \equiv \nabla^\mu \phi = g^{\mu\nu} \nabla_\nu \phi \; .
\end{equation}
Note that in the conformal transformation, i.e. ${\cal B}=0$, we have ${\cal F} = {\cal G} =0$. The difference in the Christoffel symbols between the original and transformed frames is given by
\begin{align}
\delta \Gamma^\rho_{\mu\nu} &
= \delta^\rho_{(\mu} \nabla_{\nu)} \ln {\cal A} - \frac{1}{2} \left( g_{\mu\nu} + {\cal F} \phi_\mu \phi_\nu \right) \left( g^{\rho\sigma} - {\cal G} \phi^\rho \phi^\sigma \right) \nabla_\sigma \ln {\cal A}
\nonumber\\ & \quad
+ {\cal G} \, \phi^\rho \phi_{\mu\nu} + {\cal G} \, \phi^\rho \phi_{(\mu} \nabla_{\nu)} \ln {\cal F} - \frac{{\cal F}}{2} \, \phi_\mu \phi_\nu \left( g^{\rho\sigma} - {\cal G} \phi^\rho \phi^\sigma \right) \nabla_\sigma \ln {\cal F} \; .
\label{diff_Chris}
\end{align}
The volume measures in the two frames are related as
\begin{equation}
\frac{\sqrt{-\tilde g}}{\sqrt{-g}} = {\cal A}^{(d+1)/2} \left( 1 - {\cal F} X \right)^{1/2} \; .
\label{measure_trans}
\end{equation}

Now we compute the quantity in the parentheses on the right-hand side of \eqref{act_trans_general}. Using \eqref{diff_Chris}, we obtain
\begin{align}
\tilde g^{\mu\nu} \left( \delta \Gamma^\rho_{\mu\sigma} \delta\Gamma^\sigma_{\rho\nu} - \delta \Gamma^\rho_{\mu\nu} \delta \Gamma^\sigma_{\rho\sigma} \right)
= & \frac{1}{{\cal A}} \bigg\{ 
\frac{d(d-1)}{4} \left[ \nabla_\mu \ln {\cal A} \, \nabla^\mu \ln {\cal A} - {\cal G} \left(\phi^\mu \nabla_\mu \ln {\cal A} \right)^2 \right]
\nonumber\\ & \qquad
- \frac{d-1}{2} {\cal G} \, \nabla_\mu \ln {\cal A}
\left( \frac{1}{2} X^\mu + \phi^\mu \nabla^2 \phi + X \nabla^\mu \ln {\cal F} + \phi^\mu \phi^\nu \nabla_\nu \ln {\cal F} \right)
\nonumber\\ & \qquad
+ \frac{{\cal G}^2}{2} \left( X_\mu + X \nabla_\mu \ln {\cal F} \right) \left( \frac{1}{2} X^{\mu} + \phi^\mu \nabla^2 \phi \right)
\bigg\} \; ,
\label{action_part_Chris}
\end{align}
where $X_\mu \equiv \nabla_\mu X$ and $X^\mu = \nabla^\mu X$. Combining \eqref{act_trans_general} with \eqref{measure_trans} and \eqref{action_part_Chris}, we obtain
\begin{align}
\int d^{d+1}x \sqrt{-\tilde g} \, \tilde R = &
\int d^{d+1}x \sqrt{-g} \, {\cal A}^{(d-1)/2} \left( 1 - {\cal F} X \right)^{1/2} \Bigg\{
R - {\cal G} \, \phi^\mu \phi^\nu R_{\mu\nu}
\nonumber\\ & \quad
+ \frac{d(d-1)}{4} \left[ \nabla_\mu \ln {\cal A} \, \nabla^\mu \ln {\cal A} - {\cal G} \left(\phi^\mu \nabla_\mu \ln {\cal A} \right)^2 \right]
+ \frac{{\cal G}^2}{2} \left( X_\mu + X \nabla_\mu \ln {\cal F} \right) \left( \frac{1}{2} X^{\mu} + \phi^\mu \nabla^2 \phi \right)
\nonumber\\ & \quad
- \frac{d-1}{2} {\cal G} \, \nabla_\mu \ln {\cal A}
\left( \frac{1}{2} X^\mu + \phi^\mu \nabla^2 \phi + X \nabla^\mu \ln {\cal F} + \phi^\mu \phi^\nu \nabla_\nu \ln {\cal F} \right)
\Bigg\} \; .
\label{act_trans_general_2}
\end{align}
Using the relation
\begin{equation}
\phi^\nu R_{\mu\nu} = \left( \nabla_\rho \nabla_\mu - \nabla_\mu \nabla_\rho \right) \phi^\rho \; ,
\end{equation}
we proceed \eqref{act_trans_general_2} to
\begin{align}
\int d^{d+1}x \sqrt{-\tilde g} \, \tilde R = &
\int d^{d+1}x \sqrt{-g} \, {\cal A}^{(d-1)/2} \left( 1 - {\cal F} X \right)^{1/2} 
\Bigg\{
R + {\cal G} \left[ \phi^\mu_{\; \nu} \phi^\nu_{\; \mu} - \left( \nabla^2 \phi \right)^2 \right]
\nonumber\\ & \quad
+ \frac{d(d-1)}{4} \left[ \nabla_\mu \ln {\cal A} \, \nabla^\mu \ln {\cal A} - {\cal G} \left(\phi^\mu \nabla_\mu \ln {\cal A} \right)^2 \right]
%\nonumber\\ & \quad
- \frac{d-1}{2} \, {\cal G} \, \nabla_\mu \ln {\cal A}
\left( X \nabla^\mu \ln {\cal F} + \phi^\mu \phi^\nu \nabla_\nu \ln {\cal F} \right)
\nonumber\\ & \quad
-{\cal G} \left( \frac{1}{2} X^\mu + \phi^\mu \nabla^2 \phi \right) \Big[ \left( d-1 \right) \nabla_\mu \ln {\cal A} + \nabla_\mu \ln {\cal F} \Big]
\Bigg\} \; ,
\label{EH_appendix}
\end{align}
up to total derivatives, arriving at the expression \eqref{act_trans_full}.

\subsection{Taking unitary gauge of the transformed action}
\label{sec:full-U}

We fixed one gauge degree of freedom in performing the Hamiltonian analysis in Sec.~\ref{sec:analysis}. While we simply reported the resultant action in this gauge in the main text, we summarize its derivation in this appendix subsection. By taking the unitary gauge, namely,
\begin{equation}
\phi = t \; ,
\label{fixgauge}
\end{equation}
and by decomposing the metric in the ADM manner, i.e.
\begin{equation}
ds^2 = - N^2 dt^2 + \gamma_{ij} \left( dx^i + N^i dt \right) \left( dx^j + N^j dt \right) \; ,
\end{equation} 
where $N$ is the lapse, $N^i$ the shift, and $\gamma_{ij}$ the $d$-dimensional spatial metric. In this gauge we have the relations
\begin{equation}
\phi_\mu = -\frac{1}{N} \, n_\mu \; , \quad \phi^\mu = - \frac{1}{N} \, n^\mu \; , \quad X = - \phi_\mu \phi^\mu = \frac{1}{N^2}
\end{equation}
where $n^\mu = \left( 1/N , - N^i / N \right)$ and $n_\mu = \left(-N , 0 \right)$. The Ricci scalar can be written as
\begin{equation}
R = R^{(d)} + K^i_{\; j} K^j_{\; i} - K^2 + 2 \nabla_\mu \left(  n^\mu \nabla_\nu n^\nu - n^\nu \nabla_\nu n^\mu \right) \; ,
\label{Ricci_scalar}
\end{equation}
where $K_{ij} \equiv \left( \dot \gamma_{ij} - D_i N_j - D_j N_i \right) / (2N)$. We note some useful relations,
\begin{equation}
\nabla_\mu n^\mu = K
\; , \quad n^\nu \nabla_\nu n^\mu = h^{\mu\nu} \partial_\nu \ln N \; , \quad
\nabla_\mu n^\nu \nabla_\nu n^\mu = K^i_{\; j} K^j_{\; i} \; ,
\end{equation}
where $h^{\mu\nu} \equiv g^{\mu\nu} + n^\mu n^\nu$. 
Then the action \eqref{EH_appendix} in the unitary gauge takes the form
\begin{align}
\int d^{d+1}x \sqrt{-\tilde g} \, \tilde R = & 
\int dt d^dx \sqrt{\gamma} \, {\cal A}^{(d-1)/2} \left( N^2 - {\cal F} \right)^{1/2}
\Bigg\{
\frac{{\cal G}}{{\cal F}} \left[ K^i_{\; j} K^j_{\; i} - K^2 - (d-1) K L - \frac{d(d-1)}{4} \, L^2 \right]
+ R^{(d)}
\nonumber\\ & \quad
+ \frac{d(d-1)}{4} h^{\mu\nu} \partial_\mu \ln {\cal A} \, \partial_\nu \ln {\cal A} + (d-1) \frac{{\cal G}}{{\cal F}} \, h^{\mu\nu} \partial_\mu \ln {\cal A} \, \partial_\nu \ln N
- \frac{d-1}{2} \, \frac{{\cal G}}{N^2} \, h^{\mu\nu} \partial_\mu \ln {\cal A} \, \partial_\nu \ln {\cal F}
\Bigg\} \; ,
\label{trans_full_unitary_1}
\end{align}
up to total derivatives. Now looking at the last term in \eqref{trans_full_unitary_1},
\begin{align}
\int dt d^dx \sqrt{\gamma} \, 
& {\cal A}^{(d-1)/2} 
\left( N^2 - {\cal F} \right)^{1/2} \, \frac{{\cal G}}{N^2} \, h^{\mu\nu} \partial_\mu \ln {\cal A} \, \partial_\nu \ln {\cal F}
\nonumber\\ &
= \int dt d^dx \sqrt{\gamma} \, \frac{{\cal A}^{(d-1)/2}}{\sqrt{N^2-{\cal F}}} \, D_i {\cal F} \, D^i \ln {\cal A}
\nonumber \\ &
= \int dt d^dx \sqrt{\gamma} \, {\cal A}^{(d-1)/2} \left( N^2 - {\cal F} \right)^{1/2} 
\left[ (d-1) D_i \ln {\cal A} \, D^i \ln {\cal A} + 2 D^2 \ln {\cal A} + 2 \frac{{\cal G}}{{\cal F}} D_i \ln {\cal A} \, D^i \ln N \right]
\end{align}
up to total derivatives, where $D_i$ is the covariant derivative associated with the spatial metric $\gamma_{ij}$. Plugging this result into \eqref{trans_full_unitary_1}, we obtain
\begin{align}
\int d^{d+1}x \sqrt{-\tilde g} \, \tilde R = & 
\int dt d^dx \sqrt{\gamma} \, {\cal A}^{(d-1)/2} \left( N^2 - {\cal F} \right)^{1/2}
\Bigg\{
\frac{{\cal G}}{{\cal F}} \left[ K^i_{\; j} K^j_{\; i} - K^2 - (d-1) K L - \frac{d(d-1)}{4} \, L^2 \right]
\nonumber\\ & \quad
+ R^{(d)} - (d-1) D^2 \ln {\cal A} - \frac{(d-1)(d-2)}{4} D_i \ln {\cal A} \, D^i \ln {\cal A}
\Bigg\} \; ,
\label{act_unitary_app}
\end{align}
giving the expression \eqref{act_trans_unitary}. Instead of taking the unitary gauge for the transformed action \eqref{EH_appendix}, as is done here, one can obtain the identical result by first taking the unitary gauge for the original Einstein-Hilbert action \eqref{EHaction} and then transforming it in this gauge, which we explicitly show in the following subsection.

\subsection{Direct derivation of unitary-gauge action}
\label{sec:unitary}

In this appendix subsection, we reverse the order of the transformation and the gauge fixing as compared to the previous subsection, in order to show that these two operations commute, as a consistency check. We first take the unitary gauge as in \eqref{fixgauge} and decompose both the metrics in the ADM manner, i.e.~as in \eqref{eqn:ADM-g} for $g_{\mu\nu}$ and \eqref{eqn:ADM-gtil} for $\tilde g_{\mu\nu}$. In this case the disformal transformation \eqref{eqn:frametr} amounts to changing the variables as in (\ref{eqn:unitary-gauge-transformation}).

Now, since the $(d+1)$-dimensional spacetime Ricci scalar can be written as \eqref{Ricci_scalar} in the tilded frame, the Einstein-Hilbert action becomes, up to total derivatives,
\begin{align}
\int d^{d+1}x \sqrt{-\tilde g} \, \tilde R^{(d + 1)}
 & = \int dt \, d^d x \, \tilde N \sqrt{\tilde \gamma} \left[ \tilde K^i_{\; j} \tilde K^j_{\; i} - \tilde K^2 + \tilde R^{(d)} \right]
\label{ricci-known}
\end{align}
The extrinsic curvatures in the two frames relate to each other through
\begin{equation}
\tilde K_{ij} = \frac{{\cal A}N}{\sqrt{{\cal A}N^2-{\cal B}}} \left( K_{ij} + \frac{1}{2} \, \gamma_{ij} \, L \right)
\end{equation}
where $K_{ij}$ and $L$ are defined in \eqref{defs_unitary}.
On the other hand, the Christoffel symbols associated with the $d$-dimensional spatial metrics $\gamma_{ij}$ and $\tilde \gamma_{ij}$ are related as 
\begin{equation}
\tilde \Gamma^k_{ij} = \Gamma^k_{ij} + \delta^k_{(i} D_{j)} \ln {\cal A} - \frac{1}{2} \, \gamma_{ij} D^k \ln {\cal A} \; ,
\end{equation}
where $D_i$ is the covariant derivative with respect to $\gamma_{ij}$, and thus the $d$-dimensional spatial Ricci scalar transforms as 
\begin{equation}
\tilde R^{(d)} = \frac{1}{{\cal A}} \left[ R^{(d)} - (d-1) D^2 \ln {\cal A} - \frac{(d-1)(d-2)}{4} D_i \ln {\cal A} \, D^i \ln {\cal A} \right] \; .
\end{equation}
With the above two formulae and (\ref{ricci-known}), the Einstein-Hilbert action reduces to
\begin{align}
\int d^{d+1}x \sqrt{-\tilde g} \, \tilde R^{(d + 1)}
%  & = \int dt \, d^d x \, \tilde N \sqrt{\tilde \gamma} \left[ \tilde K^i_{\; j} \tilde K^j_{\; i} - \tilde K^2 + \tilde R^{(d)} \right]
% \nonumber\\
& = 
\int dt \, d^d x \, N \sqrt{\gamma} \left[ A_4(t,N) \left( K^2 - K^i_{\; j}  K^j_{\; i} + (d-1) K L + \frac{d(d-1)}{4} L^2 \right) - U(t,N,\gamma) \right] \; ,
\end{align}
where $A_4$ and $U(t,N,\gamma)$ are defined in \eqref{defs_unitary}, directly reproducing the result \eqref{act_unitary_app} obtained in the previous subsection.
 This explicitly demonstrates that the transformation and the gauge fixing are indeed commutative procedures.

\section{Examples of how to make transformations derivative independent}
\label{app:example}

In this appendix, we describe with simple kinematic examples the procedure how to reduce any invertible transformations to the form \eqref{eqn:general-transformation} in Sec.~\ref{sec:general}. When a transformation involves time derivatives of variables, one needs to introduce auxiliary fields to replace them. In the following subsections, we argue that the transformation procedure with this replacement is identical to the original one whenever the transformation is regular and invertible, and this identity does not hold if it is singular, i.e. non-invertible.

\subsection{Regular transformations}
\label{appsub:regular}

As an example of regular transformations,
let us consider a two-free-particle system described by the Lagrangian
\begin{equation}
\label{ho}
 L = \frac{1}{2}\dot{Q_1}^2 + \frac{1}{2}\dot{Q_2}^2.
\end{equation}
Obviously, this system has 2 degrees of freedom.
If we consider a derivative-dependent transformation of $Q_1$ and $Q_2$ as
\begin{equation}
\label{qtrans}
q_1 = Q_1 + \epsilon \dot{Q_2}^2, \qquad q_2 = Q_2,
\end{equation}
where $\epsilon$ is a constant, we can find the inverted transformation
\begin{equation}
\label{qtrans2}
   Q_1 = q_1 -\epsilon \dot{q_2}^2,   \qquad   Q_2 = q_2.
\end{equation}
Under this transformation, the Lagrangian (\ref{ho}) is transformed to 
\begin{equation}
\label{ho2}
L' = 
\frac{1}{2} \left( \dot{q}_1 - 2 \epsilon \dot{q}_2 \ddot{q}_2 \right)^2 + \frac{1}{2}\dot{q}_2^2 \; ,
\end{equation}
and in order to do the Hamiltonian analysis, we further introduce to the Lagrangian (\ref{ho2}) an auxiliary variable 
$r$ that corresponds to $\ddot{q_2}$ by adding $s(r-\ddot{q_2})$, which is equivalent to $sr+\dot{s}\dot{q_2}$ up to total derivative. One thus has the following equivalent Lagrangian up to total derivatives
\begin{equation}
L' = 
\frac{1}{2} \left( \dot{q_1} - 2 \epsilon \dot{q}_2 r \right)^2 + \frac{1}{2}\dot{q_2}^2
+ sr + \dot{s}\dot{q_2} \; . 
\end{equation}
By the Hamiltonian analysis it 
is straightforward
to show that the number of physical degrees of freedom of this system 
is
the same as the original system (\ref{ho}).  

Alternatively, as an equivalent formulation, we can replace the transformation (\ref{qtrans}) with a derivative-\textit{independent} one.
 To do this, as we have mentioned in the Sec.~4, we introduce an auxiliary variable $R$ by adding $S(R- \dot{Q_2})$, to the original Lagrangian (\ref{ho}). One thus have the following equivalent Lagrangian
\begin{equation}
\label{ho3}
L = \frac{1}{2}\dot{Q_1}^2 
+ \frac{1}{2} R^2 
+ S (R-\dot{Q_2}).
\end{equation}
Then we can consider a transformation from the set of variables ($Q_1$, $Q_2$, $R$, $S$) to a new set of variables ($q_1$, $q_2$, $r$, $s$)  instead of the transformation (\ref{qtrans}) as
\begin{equation}
 q_1 = Q_1 + \epsilon R^2, \qquad q_2 = Q_2, \qquad r = R, \qquad s = S.
\end{equation}
It is clear
that this is a point transformation, and thus regular and invertible, as well as the original transformation (\ref{qtrans}). The Lagrangian (\ref{ho3}) is rewritten with the new variables
as
\begin{equation}
 L'' = \frac{1}{2}(\dot{q}_1-2\epsilon r\dot{r})^2 
  + \frac{1}{2} r^2 
 + s (r- \dot{q_2}).
\end{equation}
By performing the Hamiltonian analysis on this Lagrangian we find that the number of degrees of freedom is again the same as the original system (\ref{ho}). For completeness, in this case the hidden constraint can be found to be $\pi_r+2\epsilon r \pi_{q_1}=0$, where $\pi_r$ and $\pi_{q_1}$ are the canonical conjugate momentum of $r$ and $q_1$ respectively.

When we consider a more complex transformation including derivatives of arbitrary orders, we can always perform the same procedure as we have shown above, namely that we could reduce it to a derivative-independent system with more auxiliary fields, as long as the transformation is regular and invertible. However, if the transformation is singular and non-invertible, our procedure may no longer hold and consequently there may be a change in the number of degrees of freedom and the evolution of a system. The reason is that the singular nature does not allow us to replace such a transformation with a new, derivative-independent one by use of auxiliary variables in a consistent manner under a straightforward application of our procedure (at least in some cases). In the following subsection, we will show an example of singular transformations to which our procedure cannot directly be applied.

\subsection{Singular transformations}

We consider another simple example described by the Lagrangian as
\begin{equation}
\label{un}
L = \frac{1}{2}\dot{Q}^2 + \frac{1}{2}Q^2.
\end{equation}
It is obvious that the number of dynamical degrees of freedom is 1. Now we transform this Lagrangian as
\begin{equation}
\label{sin}
q = \dot{Q} + Q
\end{equation}
leading to
\begin{equation}
\label{q2}
L' = \frac{1}{2}q^2
\end{equation}
up to total derivatives. This is obviously a system with 0 degrees of freedom. The transformation (\ref{sin}) is not invertible and is singular in the sense that original variable $Q$ cannot be expressed solely by the new one $q$. Consequently the time evolution of $Q$ derived from \eqref{un} cannot be obtained by that of $q$ in \eqref{q2}, and vise versa. We show below that our procedure in the previous subsection cannot be applied to this singular transformation. 

Let us start from an equivalent Lagrangian to (\ref{un})
\begin{equation}
\label{ }
L = 
\frac{1}{2} R^2 + \frac{1}{2}Q^2 + S(R - \dot{Q})
\end{equation}
and notice that this system has 1 degree of freedom as in the original one \eqref{un}. Now let us try transforming the set of variables $(Q, R, S)$ to a new one $(q, r, s)$ in the manner similar to the previous subsection as
\begin{equation}
\label{aux}
q = R + Q, \qquad r= R, \qquad s= S,
\end{equation}
which is free from derivatives and invertible, leading to 
\begin{equation}
\label{lag-invertible}
L' = 
\frac{1}{2} \, r^2 + \frac{1}{2} \left( q-r \right)^2 + s(r+\dot{r}-\dot{q}) \; .
\end{equation}
By performing the Hamiltonian analysis, one can see that the system \eqref{lag-invertible} obtained after the transformation (\ref{aux}) stays to have $1$ degree of freedom, which is the same as the original system \eqref{un}. This is in sharp contrast to the fact that the given transformation (\ref{sin}) changes the number of physical degrees of freedom from $1$ to $0$. In the above procedure we have inverted the relation \eqref{aux} to obtain \eqref{lag-invertible}. On the other hand, such inversion is not possible in \eqref{sin}, indicating that the two transformations are not equivalent, and therefore the Lagrangians \eqref{un} and \eqref{lag-invertible} represent two different systems. We cannot make use of auxiliary variables to eliminate derivatives in the transformation of this particular type, and this is due to the singular nature of the considered transformation \eqref{sin}.
%%%%%%%%%%%%%%%%%%%%%%%%%%%%%%%%%%%%%%%%%%%%%%%%%%%%%%%%%%%%%%
%%%%%%%%%%%%%%%%%%%%%%%%%%%%%%%%%%%%%%%%%%%%%%%%%%%%%%%%%%%%%%
\bibliographystyle{apsrmp}
%\bibliography{rmp-sample}

\end{document}